\documentclass[preprint,sort&compress,11pt,3p]{elsarticle}

\usepackage{moreverb}
\usepackage{mathtools}
\usepackage{amsmath}
\usepackage{algorithmic}
\usepackage{graphics}
\usepackage{graphicx}
\usepackage{caption}
\usepackage{enumitem}
\usepackage{color}
\usepackage{framed}
\usepackage{wrapfig}
\usepackage{bm}
\usepackage{mathrsfs}
\usepackage{mathabx}
\usepackage{multirow}
\usepackage{longtable}
\usepackage{subfigure}
\usepackage{hyperref}
\usepackage{paralist}
\usepackage{indentfirst}
\usepackage{relsize}
\usepackage{extarrows}
\usepackage{lineno,xcolor}
\usepackage{upgreek}
\usepackage{bm}
\usepackage{mwe}
\usepackage{xcolor}
\usepackage{booktabs}
\usepackage[flushleft]{threeparttable}
\usepackage[linesnumbered,lined,ruled]{algorithm2e}

\graphicspath{ {././} }

\newcommand{\eref}[1]{(\ref{#1})}

\hypersetup{
bookmarks=true,
bookmarksopen=true,
bookmarksnumbered=true,
unicode=false,
pdftoolbar=true,
pdfmenubar=true,
pdffitwindow=false,
pdfstartview={FitH},
pdftitle={My title},
pdfauthor={Author},
pdfsubject={Subject},
pdfcreator={Creator},
pdfproducer={Producer},
pdfkeywords={keywords},
pdfnewwindow=true,
colorlinks=true,
linkcolor=blue,
citecolor=blue,
filecolor=magenta,
urlcolor=blue
}

\journal{Computational Materials Science}

\DeclareMathOperator*{\argmin}{argmin}

\begin{document}
	
\title{\Large Three-dimensional convolutional neural network (3D-CNN) for heterogeneous material homogenization}

\author[NU1]{Chengping~Rao}
\ead{rao.che@husky.neu.edu}
\author[NU1]{Yang Liu\corref{cor}}
\ead{yang1.liu@northeastern.edu}

\cortext[cor]{Corresponding author. Tel: +1 617-373-8560}

\address[NU1]{Department of Mechanical and Industrial Engineering, Northeastern University, Boston, MA 02115, USA}

\begin{abstract}
	\small
	Homogenization is a technique commonly used in multiscale computational science and engineering for predicting collective response of heterogeneous materials and extracting effective mechanical properties. In this paper, a three-dimensional deep convolutional neural network (3D-CNN) is proposed to predict the effective material properties for representative volume elements (RVEs) with random spherical inclusions. The high-fidelity dataset generated by a computational homogenization approach is used for training the 3D-CNN models. The inference results of the trained networks on unseen data indicate that the network is capable of capturing the microstructural features of RVEs and produces an accurate prediction of effective stiffness and Poisson's ratio. The benefits of the 3D-CNN over conventional finite-element-based homogenization with regard to computational efficiency, uncertainty quantification and model's transferability are discussed in sequence. We find the salient features of the 3D-CNN approach make it a potentially suitable alternative for facilitating material design with fast product design iteration and efficient uncertainty quantification.
\end{abstract}

\maketitle

\section{Introduction}\label{sintro}

The last few decades have seen tremendous applications of heterogeneous materials in automotive industry, civil, aerospace and mechanical engineering. These materials possess superior mechanical properties attributed to the unique architecture and complex microstructure. Most common among these materials are concrete, alloys, polymers, reinforced composites, etc. A primary assumption generally made for computational modeling of composite materials is that these materials are periodic in microscope scale and the periodic microstructures can be approximated by representative elements (RVEs). To develop composite materials with unusual combination of properties, it is crucial to understand the effects of various characteristics of RVE (microstructure, constituent phase, volume fraction, etc.) on the macroscopic material properties. 

For most of composite design problems, effective material properties are used instead of taking all the constituents and microstructure into consideration. A lot of efforts have been devoted to developing mathematical and/or numerical approaches for calculating the effective/homogenized material properties. The homogenization theory, which was originally developed to study partial differential equations (PDEs) with rapidly oscillating coefficients \cite{hornung2012homogenization}, have been widely used to describe the mechanics of periodic microstructure of composites. Numerous homogenization approaches have been developed to calculate effective properties which can subsequently be used for macroscopic structural analysis. These approaches can be classified into three categories \cite{aboudi2012}: 
(1) Analytical methods, e.g., the Voigt and Reuss model \cite{voigt, reuss1929};
(2) Semi-analytical methods, e.g., generalized method of cells (GMC) \cite{aboudi2004gmc}, self-consistent scheme (SCS) \cite{scs1968, scs1978}, Mori-Tanaka method \cite{mori1973};
(3) Numerical methods, e.g., finite flement (FE) \cite{feyel1999fe2, feyel2000fe2, feyel2003fe2, miehe2002fe, smit1998fe, terada2001fe}, boundary element (BE) \cite{kaminski1999bem, okada2001bem}, fast fourier transforms (FFT) \cite{lee2011fft, eisenlohr2013fft}. Each of the aforementioned approaches has its pros and cons. For example, the Voigt and Reuss model provides a quick but rough upper and lower bounds for various properties of a heterogeneous material; however, the gap of the bounds grows with regard to the volume fraction (VF) of inclusions and degree of phase contrast \cite{kanoute2009review}. Although the numerical methods involve complicated discretizations and expensive computations, they offer a possibility to deal with homogenization of materials with arbitrary microstructures and constitutive models. These methods have been shown to be effective to model multiscale material behavior in both linear \cite{kaminski1999bem, terada2001fe, yuan2008homo} and nonlinear \cite{miehe2002fe, feyel1999fe2, feyel2003fe2, feyel2000fe2, yuan2008homo, hain2008numerical, liu2014regularized, liu2016nonlocal} problems given the properly defined material constituents and microstructure. However, when it comes to iterative computational design of composites with desired properties, these numerical approaches are not suitable owing to the huge computational cost \cite{fritzen2018two} and high-dimensional sample space \cite{olson1997computational, lookman2019active, fujii2001composite}.

With recent prevalence of data science, many machine learning (ML) approaches are applied to material modeling, analysis and design. A novel framework named materials knowledge systems (MKS) \cite{landi2010mks, fast2011mks, kalidindi2015mks} was formulated to exploit the merits of both analytical and numerical approaches. MKS has its theoretical rooted in statistical continuum mechanics theory \cite{kroner1986statistical} in which the structure-property linkage of the material is expressed as a polynomial series sum. Each term of the series is a product of local microstructure-related statistics and their corresponding physics-related (or influence) coefficients \cite{landi2010mks} which reflects the underlying knowledge of the localization relationship. The core of MKS is employing discrete Fourier transform (DFT) to calibrate these coefficients to the results obtained from finite element analysis (FEA). This framework is characterized with computational efficiency, data-driven property and remarkable accuracy in a variety of works \cite{landi2010mks, fast2011mks, kalidindi2015mks, yabansu2014mks, gupta2015mks}. There are also some other applications of ML approaches on computational materials and mechanics. Fritzen and Kunc \cite{fritzen2018two} proposed a two-stage data-driven homogenization approach for nonlinear solids. Lookman \emph{et al.} \cite{lookman2019active} employed an active learning approach to navigate the search space for identifying the candidates for guiding experiments or computations. The surrogate model and utility function are used for selecting among the unexplored data.

Traditional ML techniques rely largely on the feature engineering which is time-consuming and requires expert knowledge \cite{lecun2015deep}. Deep learning (DL) approaches have been developed to address this problem. Typical DL approaches, such as fully connected neural networks (FC-NN), convolutional neural network (CNN) and long short-term memory (LSTM), can automatically find the most salient features to be learned. These approaches have demonstrated tremendous success in a variety of applications such as speech recognition, computer vision (CV), natural language processing (NLP), etc. They turned out to excel at discovering the intricate structures within high-dimensional data \cite{lecun2015deep}. Some of the recent applications of DL approaches on material science include material classification \cite{zheng2016, Bell2015}, defect classification \cite{masci2012, cha2017deep, faghih2016deep}, microstructure identification \cite{azimi2018, chowdhury2016image}, microstructure reconstruction \cite{li2018transfer, li2018GAN}, composite strength prediction \cite{yeh1998modeling}, etc. In this paper, we are mostly concerned with the works employing DL to address multiscale problems of composites, particularly in the context of homogenization. For example, Lu \emph{et al.} \cite{lu2018data} adopted neural networks (NN) to establish a surrogate model for electric conduction homogenization. By substituting the RVE calculations with the data-driven model in multiscale modeling, a drastic saving of computational cost (of the order of $10^4$) was achieved compared with the FE$^2$ method \cite{feyel1999fe2}. Le \emph{et al.} \cite{le2015computational} proposed a decoupled computational homogenization approach for nonlinear elastic materials using NN to approximate the effective potential. Li \emph{et al.} \cite{li2018transfer} employed the transfer learning idea on CNN for microstructure reconstruction. Bhattacharjee and Matou\v{s} \cite{bhattacharjee2016nonlinear} performed both homogenization and localization on heterogeneous hyperelastic materials using a digital database and the manifold-based nonlinear reduced order model (MNROM). The mapping between the macroscopic loading conditions and the reduced space are realized through NN. Yang \emph{et al.} \cite{yang18gan} applied generative adversarial networks (GAN) to generate microstructures with desired material properties. Cang \emph{et al.} \cite{cang2017microstructure} implemented convolutional deep belief network (CDBN) to automate a two-way conversion between microstructures and their lower-dimensional feature representations. Bostanabad \emph{et al.} \cite{bost2016} adopted a supervised learning approach to characterize and reconstruct the stochastic microstructure.

Most of the above studies are image-based and perform representation learning within a 2D space. To fully capture the salient features of the microstructure, the 3D geometry should be considered. Very recently, Yang \emph{et al.} \cite{yang2018dl} showed the potential of three-dimensional CNN (3D-CNN) for effective elastic modulus homogenization for composites and demonstrated its advantages over traditional sophisticated physics-inspired approaches. In this work, we leverage the capability of 3D-CNN and design a network architecture for predicting the effective material properties of composites with complex heterogeneous microstructure. In particular, we consider the composite material whose microstructure can be modeled as a two-phase (matrix/inclusion) representative volume element (RVE) with randomly distributed inclusions. A diverse group of RVEs, or virtual experiment samples, have been created with different inclusion VFs and spatial distributions, so that the sample space is large enough to include the intrinsic features of the material. Finite element analysis is then performed for each of the samples to obtain the effective moduli through linear homogenization. The geometric information of the RVEs have been pre-processed to a structured (Euclidean) grid that the 3D-CNN can accept. The networks are then trained, verified and tested on synthetic data. The salient features of the proposed 3D-CNN approach include: (1) It provides an end-to-end solution for predicting the effective material properties of the composites with high efficiency and good accuracy given the geometric information of the corresponding RVEs; (2) It is able to reproduce the probability distribution of the material properties for the input characterized with uncertainty; and (3) Its transferability makes it extremely convenient while adding supplementary data or training a model for new datasets that come from different microstructure configurations. It is worth noticing that the proposed 3D-CNN approach is more advantageous for heterogeneous materials with multiple constituents and extremely complex microstructure since it has demonstrated extraordinary ability in handling high-dimensional inputs \cite{ji3DCNN, maturana3dCNN, kamnitsas3dcnn, yang2018dl}. 

The rest of the paper is organized as follows. Section \ref{method} describes the proposed methodology. Specifically, generation of the training dataset (based on 2000 RVEs) is presented in Section \ref{pre}. Some pre-processing procedures including the conversion of the raw data into the input format of the 3D-CNN model, computational homogenization approach to obtain the labels and rescaling of the labels are given. In Section \ref{3dcnnintro}, the basic concepts and mathematical operations involved in the 3D-CNN are briefly introduce. Section \ref{results} presents the numerical results. We first conduct a series of parametric tests on the hyperparameters of the 3D-CNN to find an optimal network architecture. Then a comparison between the 3D-CNN prediction and FEA result is made with regard to the accuracy and efficiency in Section \ref{discussion3D-CNN}. The benefits of the 3D-CNN approach over traditional FEM are discussed. The uncertainty quantification (UQ) is conducted in Section \ref{UQ} to evaluate the performance of current 3D-CNN model on the input with uncertainty. In Section \ref{transferlearning}, the transferability of the proposed 3D-CNN model to a dataset representing a different type of composite microstructure is investigated. Section \ref{conc} is devoted to conclusions of the paper and the outlook of future work.  

\section{Methodologies}\label{method}

\subsection{Generation of dataset and preprocessing}\label{pre}

\begin{figure}[t!]
	\centering
	\includegraphics[width=1.0\textwidth]{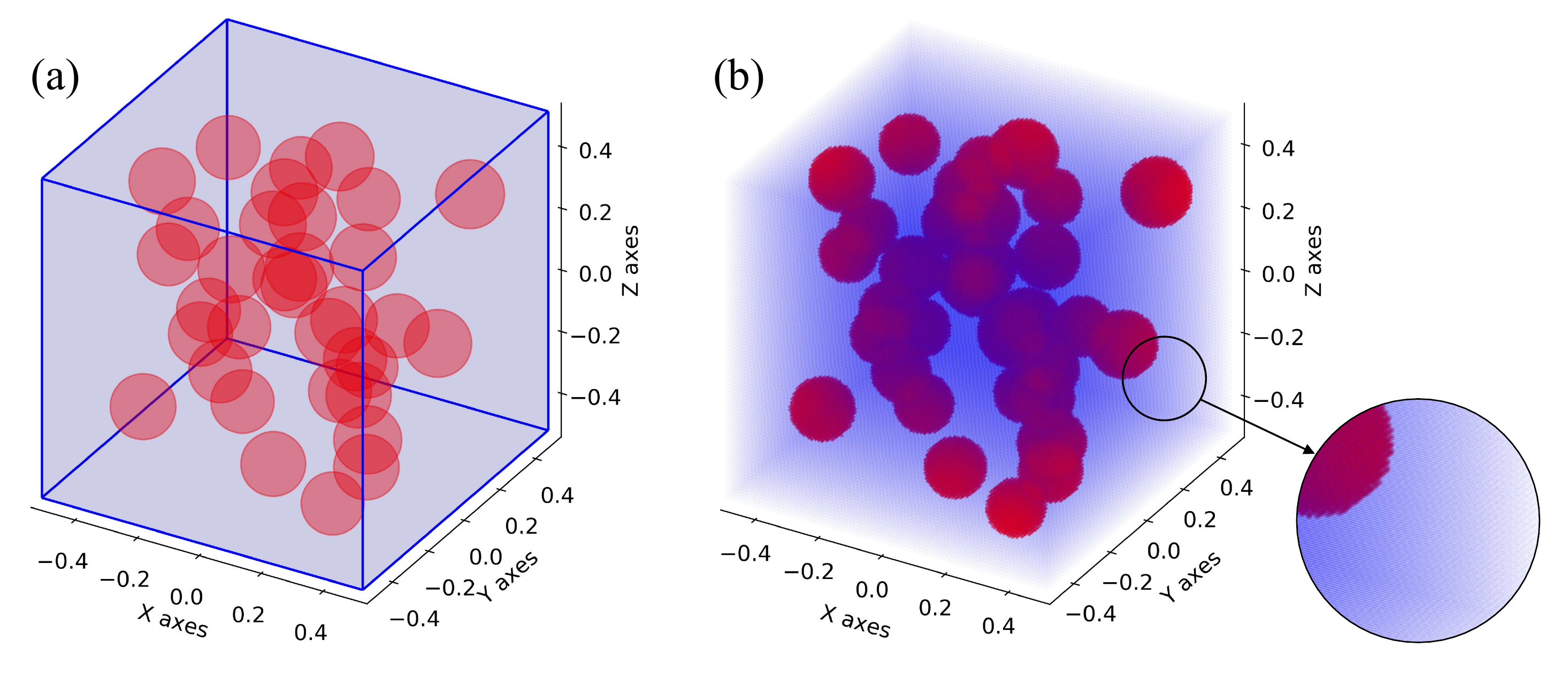}
	\caption{(a) Geometry of RVE and (b) Generated phase voxel (point cloud).}
	\label{cloud_pt}
\end{figure}

\begin{figure}[t!]
	\centering
	\includegraphics[width=0.8\textwidth]{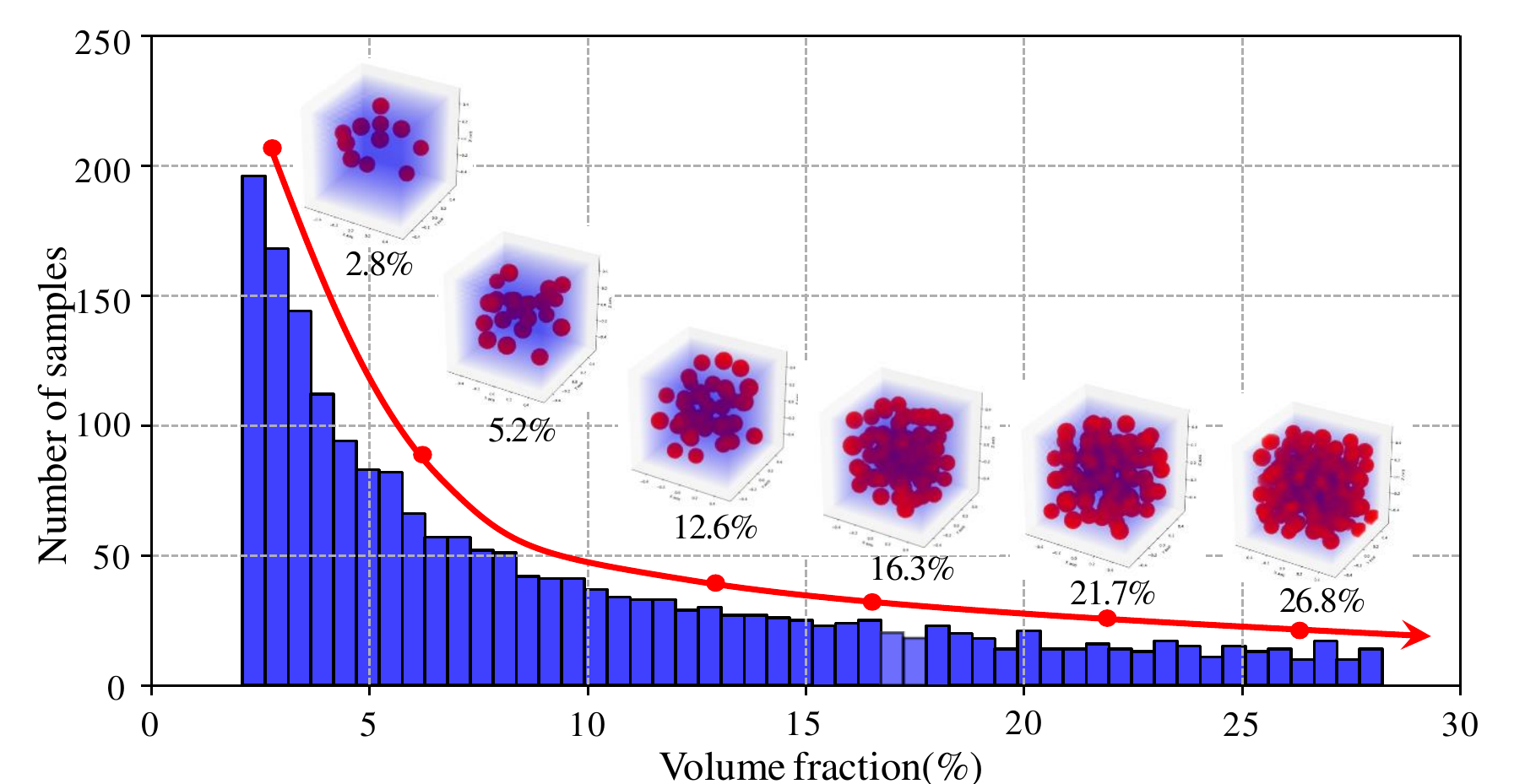}
	\caption{Distribution of number of samples with respect of the inclusion volume fraction.}
	\label{vf_distribution}
\end{figure}

In this present study, we consider particle reinforced composites, e.g., metal matrix composites, whose microstructure can be represented by a parametric two-phase RVE model with a matrix phase and a particle phase. We generate 2000 RVE samples with the volume fraction (VF) of inclusions ranging from $2\%$ to $28\%$ to establish the training data (see Fig. \ref{cloud_pt}(a)). The radius of each spherical inclusion follows a uniform distribution in the range of 0.05$\sim$0.1 mm while the length of the square RVE is 1.0 mm. The spherical inclusions within the RVE are randomly distributed based on the Hierarchical Random Sequential Adsorption (HRSA) algorithm \cite{bai2014auto} that could achieve a user-defined desired VF. Generally the RVE with low inclusion VF demonstrates greater randomness in terms of particle spatial distributions resulting in significant randomness of the effective elastic moduli. To resolve this issue, we impose an exponential distribution on the number of samples with regard to the VF, as shown in Fig. \ref{vf_distribution}, to most likely cover the manifold of the relationship between random RVEs and the effective elastic properties. This practice is meant to better capture the spatial characteristics of the RVE during training. 

Preprocessing is required to convert the geometric data (or discretized mesh data) of RVEs into Euclidean grids, the input format that a 3D-CNN can take. We resample the phase information of RVEs within fixed Cartesian grids. In particular, these RVEs are converted to $101\times101\times101$ voxels where matrix phase is denoted by 0 while inclusion phase by 1 (see Fig. \ref{cloud_pt}(b)). Given the center location and geometric information of all these inclusions, a level-set function is used to assign a binary phase value $p$ to the voxel with coordinate $(x, y, z)$, namely,

\begin{equation} 
\mathit{p}(x,y,z)= \begin{cases}
 1;  & \text{if } \sqrt{(x-x_i)^2+(y-y_i)^2+(z-z_i)^2} < r_i,~\exists~i\in{\{1,2,...n\}}\\
 0; & \text{otherwise}
\end{cases}
\label{levelset} 
\end{equation}
where $n$ is the total number of inclusions; $x_i$, $y_i$, $z_i$ and $r_i$ are coordinates of center and radius of $i$th spherical inclusion, respectively. It is noted that the size of 101 (length of 0.01 mm) is selected in order to cover all the microstructural details within the RVEs since the minimum radius for the spherical inclusion is 0.05 mm.

\begin{table}[t!]
\small
\centering
\caption{Material properties for RVE with spherical inclusions.}
\begin{tabular}{l c c}
\hline
\multicolumn{1}{l}{Materials} & \multicolumn{1}{l}{Young's modulus (GPa)} & \multicolumn{1}{l}{Poisson's ratio}   \\ \hline
Matrix     &   68.9     &   0.33      \\
Inclusion  &   379.2    &   0.21      \\ \hline
\end{tabular}
\label{matprop}
\end{table}

The deep learning method falls into the category of supervised learning in which training data needs to be labelled. In this paper, linear elastic materials are considered for both the matrix and inclusion phases. The material properties of each single phase used in this study are given in Table \ref{matprop}. Since the considered composite is assumed to be orthotropic, its constitutive tensor has 9 independent variables from which the following vector of effective material properties can be obtained:
\begin{equation}
\centering
 \mathbf{y}=\begin{bmatrix} 
E_{11}&E_{22}&E_{33}&G_{23}&G_{13}&G_{12}&\nu_{21}&\nu_{31}&\nu_{12}&\nu_{32}&\nu_{13}&\nu_{23}
\end{bmatrix}^T\label{matparams}
\end{equation}
where $\mathbf{y}$ denotes the label for each RVE sample; $E$'s, $G$'s and $\nu$'s denote the effective elastic modulus, shear modulus and Poisson's ratio, respectively, along different directions. The computational homogenization is conducted based on the framework of the classical mathematical homogenization theory \cite{guedes1990homo, yuan2008homo} via FEM. Specifically, the homogenized constitutive tensor can be calculated through averaging $\Sigma_{ij}^{mn}(\mathbf{\xi})$ over the entire volume $\Theta$ of the RVE, expressed as
\begin{equation}\label{eq_homo:1}
\overline{L}_{ijmn}=\cfrac{1}{|\Theta|}{\displaystyle\int_{\mathbf{\xi}\in\Theta}\Sigma_{ij}^{mn}(\mathbf{\xi})d\mathbf{\xi}}
\end{equation}
in which $\Sigma_{ij}^{mn}(\mathbf{\xi})$ is the stress influence function with regard to the fine-scale coordinate $\mathbf{\xi}$. It can be interpreted as the fine-scale stress induced by an unit overall strain $\epsilon_{mn}^{c}$. The implementation of numerical homogenization is achieved by solving a RVE (or unit cell) problem under periodic boundary conditions (PBCs) and unit thermal strain \cite{yuan2008homo}. The components of constitutive tensor can then be obtained by averaging the stress field over the volume, given by
\begin{equation}\label{eq_homo:2}
\overline{L}_{ijmn}=\cfrac{1}{|\Theta|}{\displaystyle\int_{\mathbf{\xi}\in\Theta}\sigma_{ij}^{mn}(\mathbf{\xi})d\mathbf{\xi}}
\end{equation}
The constitutive tensor $\underline{\underline{\mathbf{C}}}$ can be represented in the Voigt notation, written as
\begingroup
\renewcommand*{\arraystretch}{1.2}
\begin{equation}\label{c_mat}
\begin{bmatrix}
\sigma_{xx}\\ \sigma_{yy}\\ \sigma_{zz}\\ \sigma_{yz}\\ \sigma_{zx}\\ \sigma_{xy} 
\end{bmatrix}
=\underbrace{
\begin{bmatrix}
C_{11} &C_{12}  &C_{13}  &0  &0  &0 \\ 
C_{12} &C_{22}  &C_{23}  &0  &0  &0 \\ 
C_{13} &C_{23}  &C_{33}  &0  &0  &0 \\ 
0 & 0 & 0 & C_{44} & 0 & 0 \\ 
0 & 0 & 0 & 0 & C_{55} & 0 \\ 
0 & 0 & 0 & 0 & 0 & C_{66}
\end{bmatrix} }_{\underline{\underline{\mathbf{C}}}}
\begin{bmatrix}
\epsilon_{xx}\\ \epsilon_{yy}\\ \epsilon_{zz}\\ \gamma_{yz}\\ \gamma_{zx}\\ \gamma_{xy} 
\end{bmatrix} 
\end{equation}
\endgroup
The inverse of $\underline{\underline{\mathbf{C}}}$ results in the so-called stiffness matrix $\underline{\underline{\mathbf{S}}}$ shown as follows, from which the vector of effective material properties can be calculated. 
\begingroup
\renewcommand*{\arraystretch}{1.5}
\begin{equation}\label{s_mat}
\underline{\underline{\mathbf{S}}}=\underline{\underline{\mathbf{C}}}^{-1}=\begin{bmatrix}
\frac{1}{E_{11}} &-\frac{\nu_{21}}{E_{22}}  &-\frac{\nu_{31}}{E_{33}}  &0  &0  &0 \\ 
-\frac{\nu_{12}}{E_{11}} &  \frac{1}{E_{22}} &-\frac{\nu_{32}}{E_{33}}  &0  &0  &0 \\
-\frac{\nu_{13}}{E_{11}} & -\frac{\nu_{23}}{E_{22}} & \frac{1}{E_{33}}  &0  &0  &0 \\
0 & 0 & 0 & \frac{1}{G_{23}} & 0 & 0 \\ 
0 & 0 & 0 & 0 & \frac{1}{G_{31}} & 0 \\ 
0 & 0 & 0 & 0 & 0 & \frac{1}{G_{12}}
\end{bmatrix}
\end{equation}
\endgroup

The entire dataset is randomly divided into training, validation and testing set with a ratio of 1400:300:300. The training set is used for learning the parameters (i.e., weights and biases) of the 3D-CNN (see Section \ref{3dcnnintro}) while the validation set is used to tune the hyperparameters (i.e., the architecture) of the 3D-CNN. The validation set is also adopted as a regularizer via early stopping, i.e., to stop the training when the loss function on the validation set increases, as it is a sign of overfitting to the training data set \cite{ripley1996pattern}. The testing set, which is usually unseen to the training process, serves for confirming and evaluating the actual predictive power of the trained deep learning model. 

Since the RVEs in this paper are generated artificially, we can directly extract the microstructure information from the formatted data. However, how to obtain the phase information of samples from field measurements is an issue of interest. The nondestructive imaging techniques such as X-ray micro-topography \cite{stienon2009xray, proudhon2007xray, Alp2020}, 3-D atom probe \cite{kelly2007atom} and automated serial sectioning \cite{spowart2006automated} have made possible to capture 3D material microstructures. These imaging techniques are characterized with high resolution. For example, the synchrotron radiation micro-tomography is able to sample microstructure with resolution of 2048 voxels in each dimension \cite{betz2007imaging}. Therefore, it will be promising for field measurement techniques to be incorporated into current framework with appropriate down-sampling on the microstructure data. Nevertheless, this is beyond the scope of the current study.

\subsection{3D convolutional neural network}\label{3dcnnintro}

\begin{figure}[t!]
	\centering
	\includegraphics[width=1.0\textwidth]{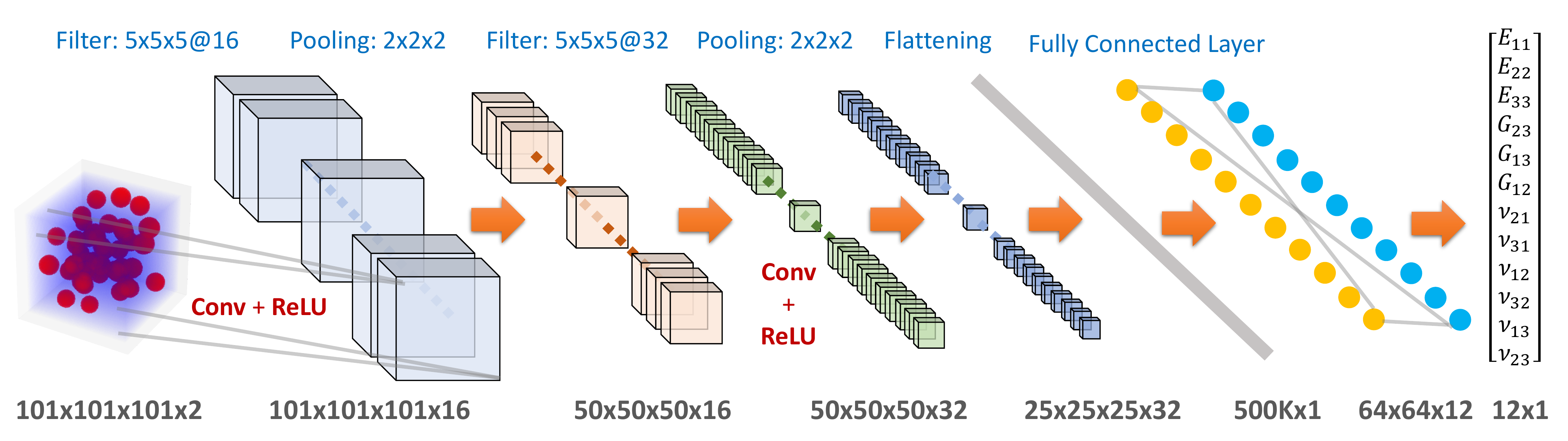}
	\caption{Proposed 3D-CNN architecture for effective properties prediction of heterogeneous materials.}
	\label{CNN_ARCH}
\end{figure}

The convolutional neural network (CNN or ConvNet) is proposed originally to solve computer vision problems. LeCun \emph{et al.} \cite{lecun1989} designed one of the very first CNNs to successfully recognize handwritten digits in 1990s. The applications of CNNs were limited by the less powerful computational ability at that time. In recent years, the CNN approach has been revived owing to the huge advancements on computational hardware such as the general purpose graphics processing units (GPUs). The CNN differs from the classical FC-NNs by its weights sharing mechanism. In this study, we propose a 3D-CNN architecture (see Fig. \ref {CNN_ARCH}) for inferring homogenized/effective material properties (e.g., elastic moduli, shear moduli and Poisson's ratio) from given microstructure configurations (e.g., discretized distribution of material phases). 

\begin{figure}[t!]
	\centering
	\includegraphics[width=1.0\textwidth]{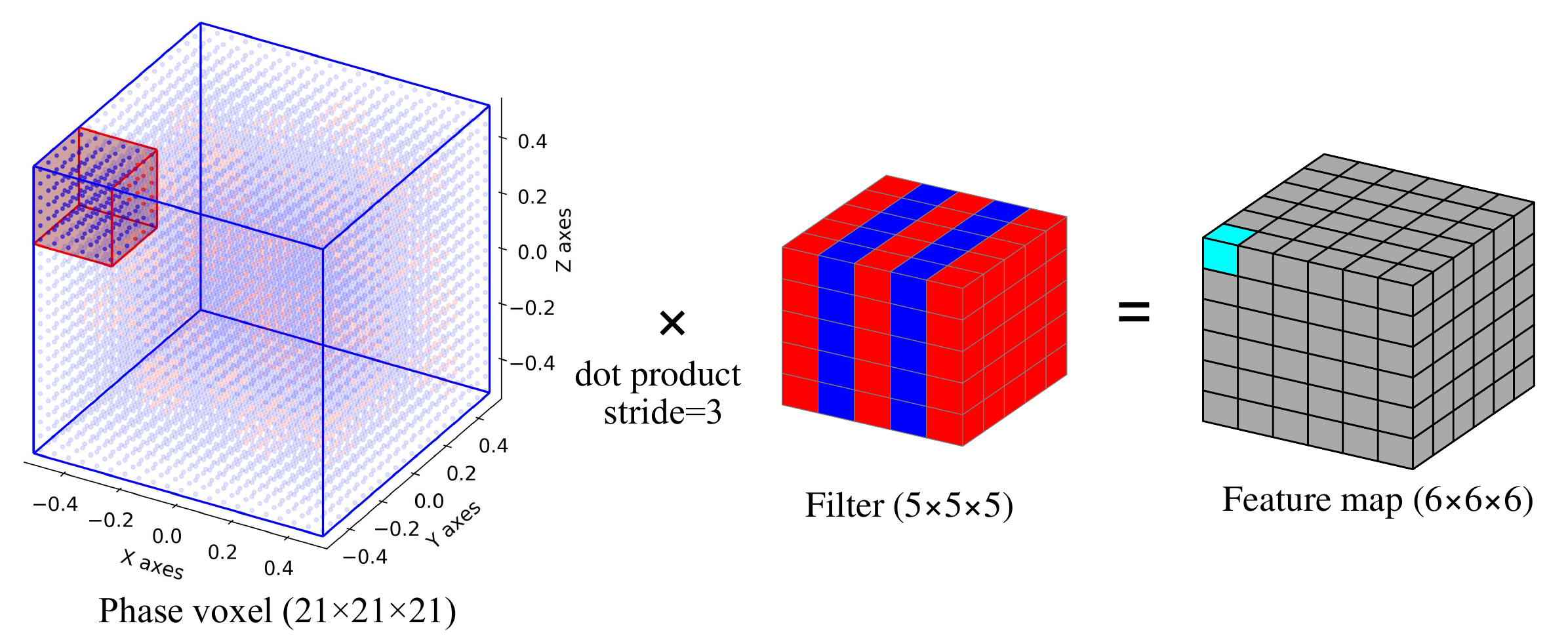}
	\caption{Convolution operation in the 3D-CNN model.}
	\label{Conv_op}
\end{figure}

\begin{figure}[t!]
	\centering
	\includegraphics[width=0.8\textwidth]{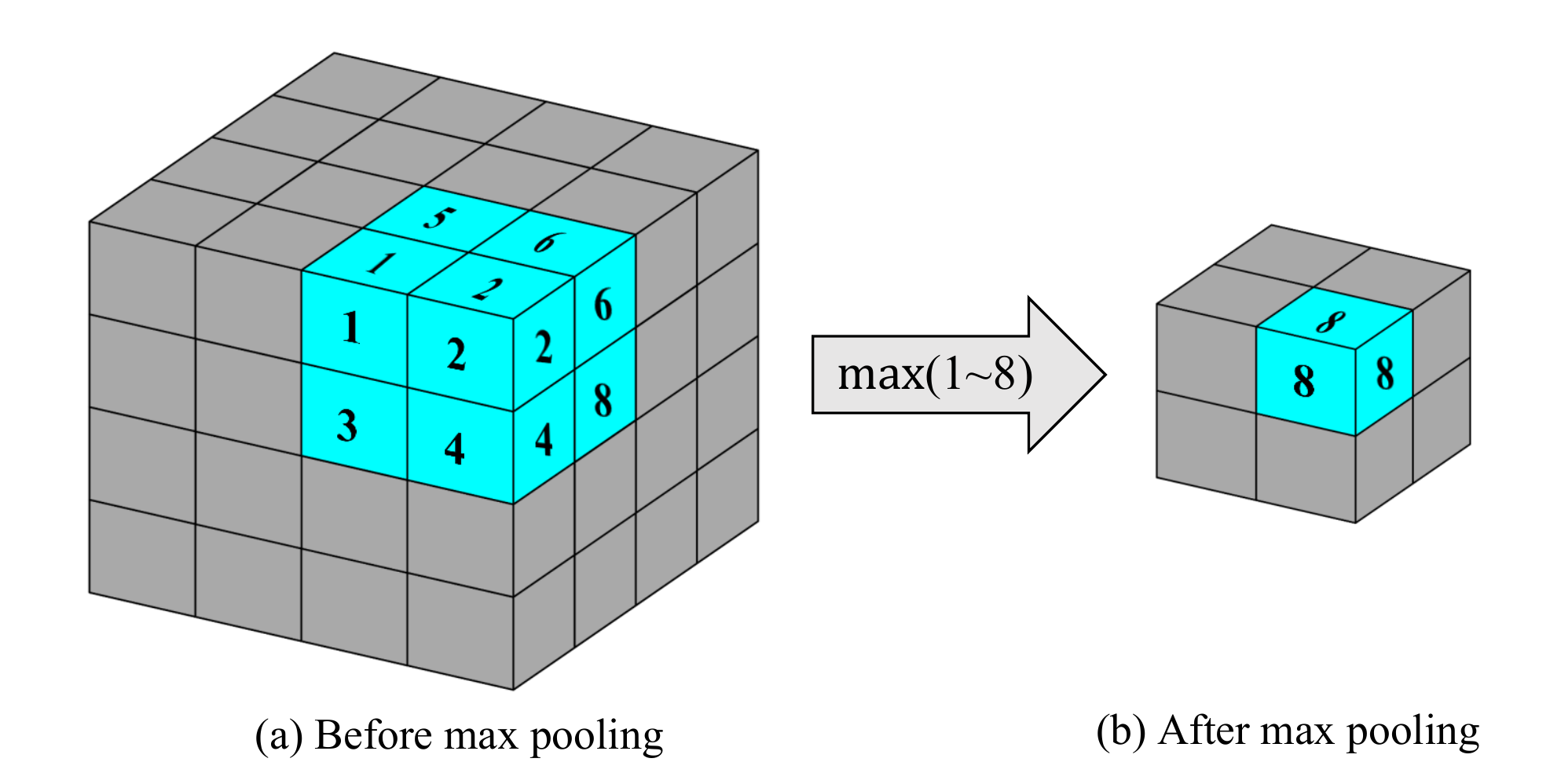}
	\caption{Max pooling operation in the 3D-CNN model.}
	\label{Pool_op}
\end{figure}

\begin{figure}[t!]
	\centering
	\includegraphics[width=0.5\textwidth]{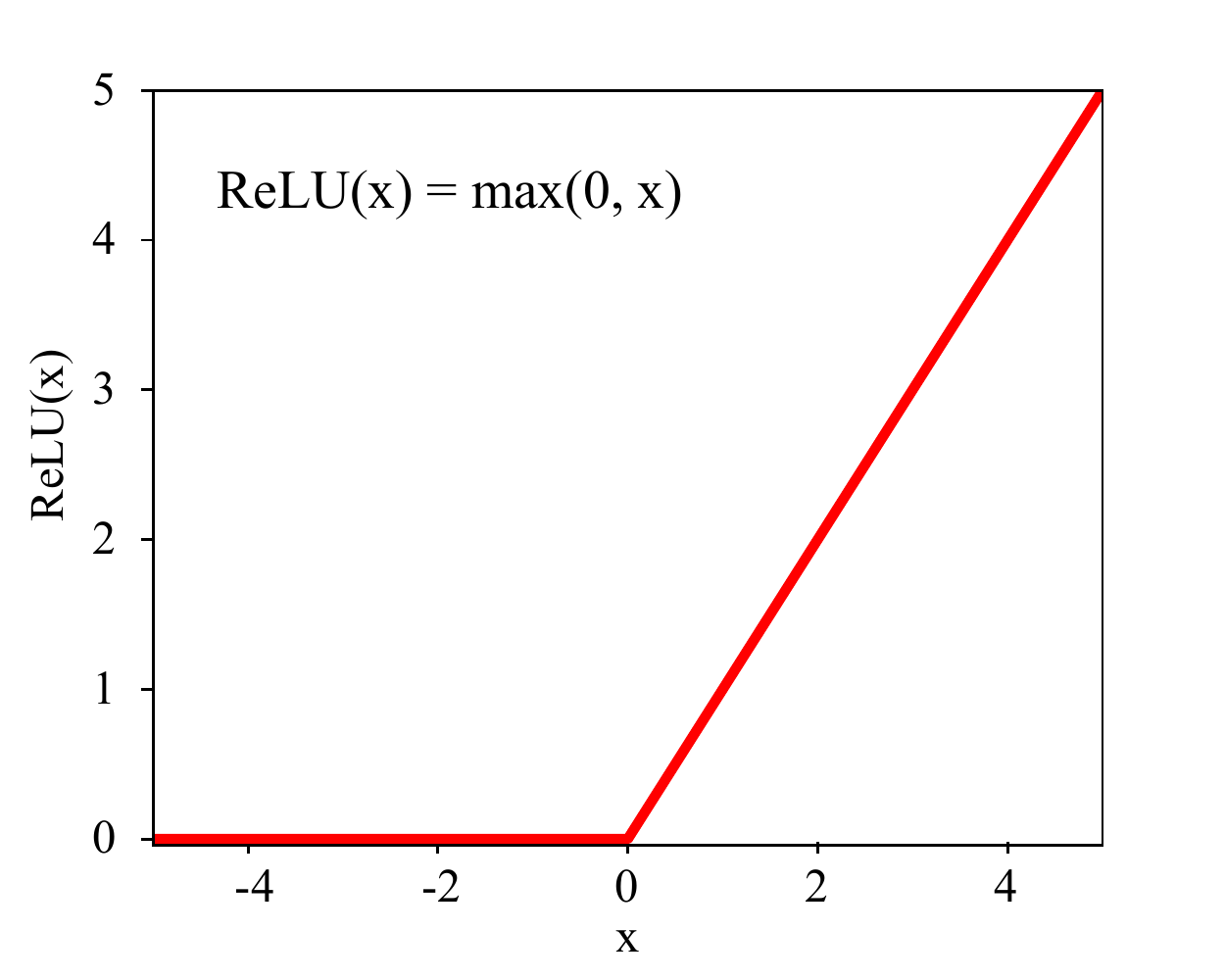}
	\caption{Rectified linear unit (ReLU) function as the activation function.}
	\label{relu}
\end{figure}

The 3D-CNN takes the preprocessed phase voxels as the input. Subsequent multiple convolutional layers serve as the critical composition of the CNN with 3D convolution filters and pooling operation. As indicated in Fig. \ref{Conv_op}, the 3D filter scans over the phase voxels and applies convolutional operation (dot product of tensor) to produce the feature map. The weights and biases of each filter are trained to extract the salient features from the input. Stride, padding and filter size are a few common hyperparameters defining convolutional operations. Stride denotes the size of step that filters move each time. For instance, the stride length of 1 means the filters scan the volume voxel by voxel. To preserve the spatial size of the output, it is convenient to pad the input with zero-value voxels. A good example is that the input and output size in Fig. \ref{Conv_op} will be identical ($21\times21\times21$) if the convolution operations are conducted with stride of 1 and 2-layer zero padding. Pooling layers are usually added between successive convolutional layers in the CNN. It progressively reduces the spatial size of data through down-sampling the voxel value. Pooling operations may compute the maximum or average value within a volume. Fig. \ref{Pool_op} demonstrates how the max-pooling operation works with volume size of $2\times2\times2$. The activation layers are employed to introduce nonlinearity into the CNN. It takes a single number and performs a certain fixed mathematical function. Some typical activation functions are Rectified Linear Unit (ReLU) $f(x)=\max(0, x)$, Sigmoid function $f(x)={\mathrm{1} }/{(\mathrm{1} + e^{-x})}$ and tanh function $f(x)=\tanh(x)$. Among these non-linear functions, ReLU (see Fig. \ref{relu}) is preferred and thus selected owing to its cheap arithmetic operation and excellent convergence properties on the stochastic gradient descent (SGD) algorithm compared with the Sigmoid or tanh functions. Mathematical expression of the output value $\gamma$ at position $(x, y, z)$ on $j$th feature map in $i$th 3D convolutional layer can be written as \cite{ji3DCNN}
\begin{equation} 
 \gamma_{j,xyz}^{(i)}={\rm ReLU}\left(b_{j}^{(i)}+\sum_{m=1}^{M^{(i-1)}}\sum_{p=0}^{P^{(i)}-1}\sum_{q=0}^{Q^{(i)}-1}\sum_{r=0}^{R^{(i)}-1}w_{jm,pqr}^{(i)}\gamma_{m,(x+p)(y+q)(z+r)}^{(i-1)} \right )
\end{equation}
where ${\rm ReLU(\cdot)}$ denotes element-wise ReLU function; $b_{j}^{(i)}$ is the common bias for $j$th feature map; $w_{jm,pqr}^{(i)}$ is the $(p, q, r)$th value of the 3D filter for $j$th feature map at $i$th layer associated with the $m$th feature map in the $(i-1)$th layer; $M^{(i-1)}$ is the number of feature maps at $(i-1)$th layer; $P^{(i)},~Q^{(i)}$ and $R^{(i)}$ denotes the size of the 3D filter at $i$th layer. In this paper, an constant filter size is used through the convolutional layers.
 
FC layers are employed at the end of the 3D-CNN where neurons between two neighboring layers are interconnected. FC layers take the flattened tensor from the previous hidden layer as the input and map them to desired output which are exactly the vector of effective material properties with length of 12 as shown in Eq. \eref{matparams}. The connection between two adjacent layers, here from $(i-1)$th to $i$th, can be expressed concisely in the form of tensor operations, given by
\begin{equation} 
 \boldsymbol{\gamma}^{(i)}={\rm \sigma}\left(\mathbf{{W}}^{(i)}\boldsymbol{\gamma}^{(i-1)} + \mathbf{b}^{(i)}\right)
\end{equation}
where $\boldsymbol{\gamma}^{(i-1)}$ and $\boldsymbol{\gamma}^{(i)}$ are the input and output for the $i$th layer; ${\rm \sigma(\cdot)}$ denotes the Sigmoid activation function acting element-wise; $\mathbf{W}^{(i)}$ and $\mathbf{b}^{(i)}$ are the weight matrix and bias vector between the $i$th and the $(i-1)$th FC layers. The weights and biases in the FC layers are also the trainable parameters of the 3D-CNN. The mean square error (MSE) between the 3D-CNN's prediction and the ground truth of the training dataset is adopted as the loss function, given by
\begin{equation} 
\begin{aligned}
\mathcal{L}(\mathbf{W},\mathbf{b}|\mathbf{D})=\frac{1}{n}\sum_{k=1}^{n}\sum_{l=1}^{12}\left(\text{y}_{kl}^{truth}-\text{y}_{kl}^{pred}\right)^2
\end{aligned}
\label{mse} 
\end{equation}
where $\mathbf{D}$ denotes the training data set \{$\mathbf{x}_k,\mathbf{y}_k$\}, $n$ denotes the total number of samples, $l$ denotes the index of component for the effective properties vector. The optimal parameters $\{\mathbf{W^*}$,$\mathbf{b^*}\}$ can be obtained by minimizing the loss function, namely,
\begin{equation} 
\{\mathbf{W^{*}},\mathbf{b^{*}}\} = \argmin_{\{\mathbf{W},\mathbf{b}\}} \left\{\mathcal{L}(\mathbf{W},\mathbf{b}|\mathbf{D})\right\}
\label{argmin} 
\end{equation}

A common issue facing the DNN-based approaches is to mitigate the overfitting brought about by its extraordinary approximation ability. Several treatments are considered in this paper. Firstly, it is noted that there is a scale difference between the outputs of elastic (or shear) modulus and Poisson's ratio which might bring problems to the optimization. For example, an output variable with a large range of values could result in large error gradient values causing weight values to change dramatically, making the learning process unstable \cite{bishop1995neural}. Therefore, label rescaling is employed here to address this problem. The elastic (or shear) moduli and Poisson's ratios are scaled separately into the range of 0 to 1 with a min-max scaling manner, e.g.,
\begin{equation} 
\bar{\mathbf{y}}=\frac{\mathbf{y}-\min(\mathbf{y})}{\max(\mathbf{y})-\min(\mathbf{y})}
\label{minmaxscale} 
\end{equation}
where $\mathbf{y}$ denotes the output component vector while $\bar{\mathbf{y}}$ is the corresponding scaled output. In addition to label rescaling, early stopping \cite{girosi1995regularization} and sample shuffling during training are adopted as the regularizer to alleviate overfitting. 

In this paper, the filter size of $5\times5\times5$, stride length of 1 and no-padding are configured on the convolutional layers. The max pooling with size $2\times2\times2$ is set on pooling layer. ReLU function is selected as the activation function due to the aforementioned merits. Other hyperparameters such as number of filters, depth of convolutional layers and FC layers are selected through parametric tests on Section \ref{parameterictest}. An adaptive learning rate optimization algorithm, Adam \cite{kingma2014adam}, is used for the training of 3D-CNN models.

\section{Results}\label{results}

In this section, the performance of the proposed 3D-CNN for heterogeneous material homogenization is evaluated. A series of parametric tests on the network hyperparameters (e.g., filter size, depth, width) of the 3D-CNN are conducted to find a suitable architecture for the current application. Then the trained 3D-CNN is used to predict the effective properties on the testing dataset with 300 RVEs. The performance of the 3D-CNN is discussed based on a comparison between the model inference and the results produced by traditional FEA. Since the randomness of the inclusion distribution is a significant aspect of the naturally occurring heterogeneous materials, uncertainty quantification is conducted on an independent dataset that imitates the input with uncertainty. Finally, the transferability of the trained 3D-CNN model to a new dataset (for RVEs with different inclusion shapes) is examined. The proposed 3D-CNN architecture is implemented with the high-level neural networks API - Keras \cite{2015keras} using Python 3.7. Our networks are trained on platform equipped with NVIDIA GeForce GTX 1080 Ti GPU and Intel Core i9-7980XE CPU@2.60GHz. 

\subsection{Design of the 3D-CNN architecture}\label{parameterictest}

A typical CNN involves dozens of hyperparameters that control the learning process of the network. These include the number of filters, filter size, learning rate, number of hidden layers, and batch size, just to name a few. The huge sample space makes it nearly impossible to find an optimal combination of hyperparameters. Therefore, the hyperparameters are usually searched in a trail-and-error manner within a small sample space. Fortunately, some rules of thumb for selecting the hyperparameters can be applied here. For example, the number of filters in convolutional layer should reflect the enrichment of characterized features within the input. It usually depends on the number of samples and the complexity of the task \cite{krizhevsky2012imagenet}. The number of FC layers and neurons determine directly the total number of parameters (weights and biases) and thus affect the representational power of the network \cite{cybenko1989approximation}. Therefore, it is natural to select the hyperparameter combination based on the underlying physical and mathematical interpretation of the ``knowledge'' to be learned. In this section, We evaluate different 3D-CNN architectures with varying number of hidden layers and filters. The MSE on validation dataset) defined in Eq. \eref{mse} is used to measure the performance of each 3D-CNN architecture.

As is mentioned in Section \ref{pre}, the Cartesian grid used to sample the RVE is of size $101\times101\times101$ so that the smallest inclusion with radius equaling 0.05 mm could be captured. In our design of the 3D-CNN architecture, we select the fixed filter size to be 5 in all three dimensions so that it is identical to the size of smallest inclusion. The batch size during training is set to be 25 according to the memory space available on the hardware. The trained model with the best performance, i.e., lowest MSE, for each architecture after 1000 epochs are saved for later inference. This is the commonly used technique aforementioned as early stopping. Table \ref{performance} provides the configurations of each 3D-CNN architecture. The convolutional layer and fully connected layer are denoted by Conv($\cdot$) and FC($\cdot$) respectively. The values within the bracket of Conv($\cdot$) indicates the filter number and filter size. Similarly the values within the bracket of FC($\cdot$) represent the number of neurons (width) in each layer. For example, Conv(32, 5) means the convolutional layer has 32 filters whose size is $5\times5\times5$ while FC(64, 32) means the FC layers are composed of two layers whose widths are 64 and 32 respectively. 

The corresponding MSE of each architecture is listed on Table \ref{performance}. It can be inferred from Case 2 and Case 5 that increasing the number of filters in each convolutional layer does not necessarily improve the prediction performance. A large width of the network might cause the overfitting issue on the training dataset. A similar situation is met while increasing the number of convolutional layers (e.g., Case 4) and FC layers (e.g., Case 7) on the basis of Case 2. Moreover, the comparison between Cases 1-3 demonstrates that two FC layers each with 64 and 32 neurons deliver the best prediction performance on unseen validation dataset. Taking both the accuracy and efficiency of the listed architecture into account, the 3D-CNN architecture with hyperparameters shown in Case 2 is employed in the remaining of this paper.

\begin{figure}[t!]
	\centering
	\includegraphics[width=1.0\textwidth]{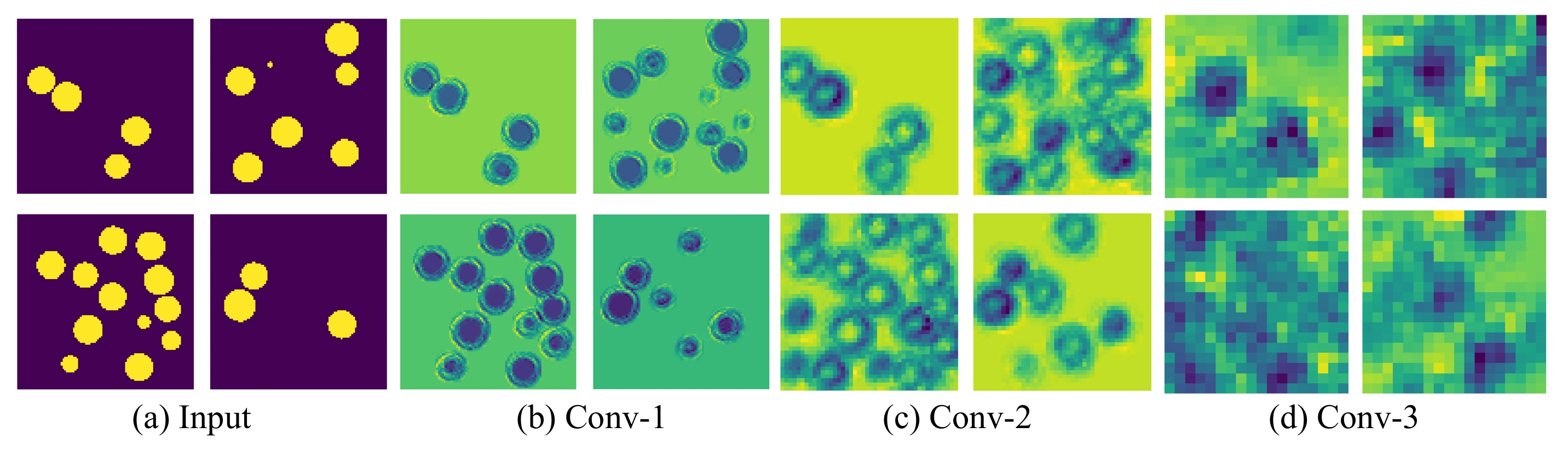}
	\caption{Visualization of the input slice and the feature map slice.}
	\label{featuremap}
\end{figure}

To check how the material phase information (input) is transformed through the multiple convolution layers, a group of example feature map slices are visualized in Fig. \ref{featuremap}. The feature map is 3D in the present 3D-CNN approach. However, for easier visualization, we only show the slices of the feature map. Typically the colored area in the feature map is called activated region which represents the extracted feature from the input. In our application, the activated region reflects the microstructural characteristics that the convolutional filters capture. It can be seen that the first convolutional layer preserves most of the details in the original input. As we go deeper into the convolutional layer, the feature map becomes abstract because it usually represents the high-level characteristics that is less visually recognizable. 

\begin{table}[t!]
\caption{Performance comparison of various 3D-CNN architecture.}
\small
\centering
\begin{tabular}{l p{12cm} l l l}
    \hline
    \multicolumn{1}{l}{No.} & \multicolumn{1}{l}{Model description} & \multicolumn{1}{l}{MSE}\\ 
    \hline
    1   & Conv(16,5)+Conv(16,5)+Conv(32,5)+FC(32$\times$ 16)  &   2.82$\times$ $10^{-4}$    \\
    2   & Conv(16,5)+Conv(16,5)+Conv(32,5)+FC(64$\times$ 32)   &  2.79$\times$ $10^{-4}$      \\
    3   & Conv(16,5)+Conv(16,5)+Conv(32,5)+FC(128$\times$ 64)  &   2.89$\times$ $10^{-4}$    \\
    4   & Conv(16,5)+Conv(16,5)+Conv(16,5)+Conv(32,5)+FC(64$\times$ 32)  &   6.33$\times$ $10^{-4}$    \\
    5   & Conv(16,5)+Conv(32,5)+Conv(32,5)+FC(64$\times$ 32)   &  3.61$\times$ $10^{-4}$     \\
    6   & Conv(16,5)+Conv(16,5)+Conv(16,5)+FC(64$\times$ 32)  &   3.44$\times$ $10^{-4}$    \\
    7   & Conv(16,5)+Conv(16,5)+Conv(32,5)+FC(64$\times$ 32$\times$ 32)  &  2.86$\times$ $10^{-4}$     \\
    \hline
\end{tabular}
\label{performance}
\end{table}

\subsection{Prediction of effective properties} \label{discussion3D-CNN}

\begin{figure}[t!]
	\centering
	\includegraphics[width=1.0\textwidth]{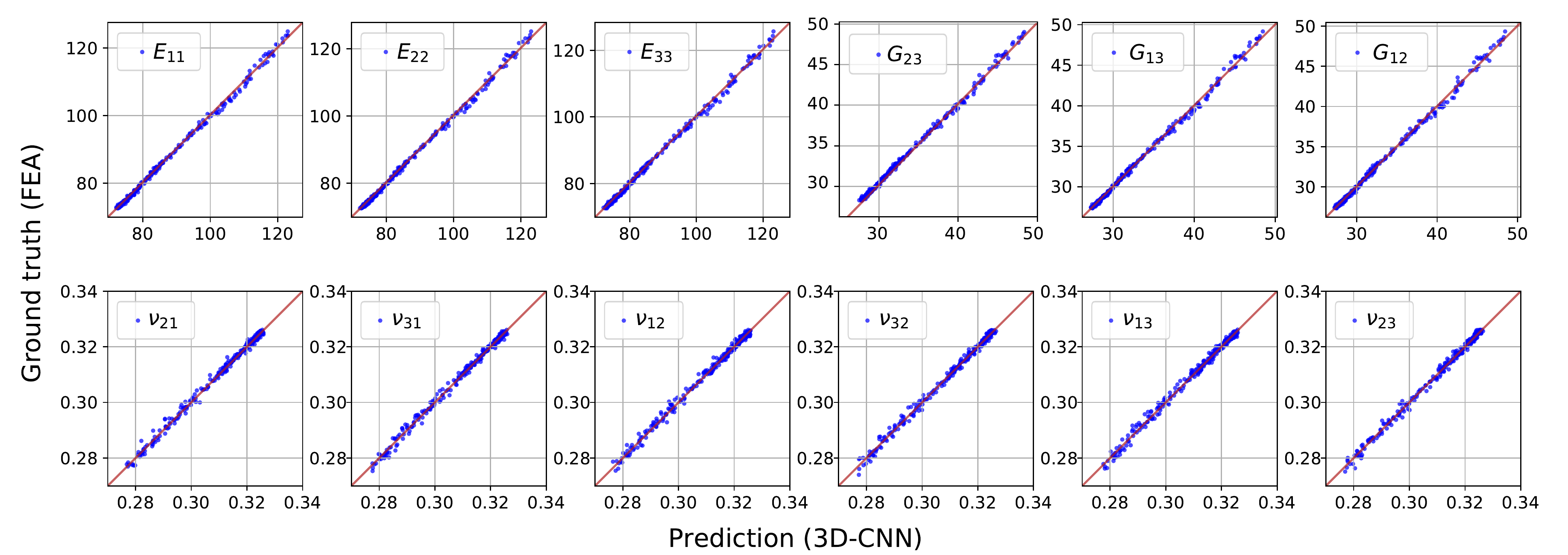}
	\caption{Comparison between the 3D-CNN prediction and ground truth (FEA).}
	\label{predvstruth}
\end{figure}

In this part, the performance of the trained 3D-CNN model is evaluated on the validation dataset which consists of 300 RVEs with the same VF range (e.g., 2\%-28\%). The prediction and ground truth (obtained through FEA) for the effective properties of each RVE sample is shown as scatter plots in Fig. \ref{predvstruth}. Since the baseline is given as red line, we can see that the trained model gives accurate prediction for the 12 components of Young's modulus, shear modulus and Poisson's ratio. It is also observed that the prediction on the samples with low VFs, e.g., the left-bottom part of the scatter plots for moduli ($E$'s and $G$'s) and the upper-right part for Poisson's ratios ($\nu$'s), perform identically well as the counterpart with high VFs, though larger randomness is present for RVEs with low VFs. Let us recall that, in Section \ref{pre}, an exponential distribution of sample number against VF is imposed while generating the datasets. As a result, the number of low VF samples is much greater than the number of the high VF samples, which alleviates the issue of low VF induced uncertainty. To measure the prediction performance quantitatively, we calculate the mean absolute relative error (MARE) for each component, defined as
\begin{equation} 
\begin{aligned}
\text{MARE}=\frac{1}{n} \sum_{i=1}^{n} \frac{|\hat{\text{y}}_i-\text{y}_i|}{|\text{y}_i|}
\end{aligned}
\label{mare} 
\end{equation}
where $\hat{\text{y}}_i$ and $\text{y}_i$ are prediction and ground truth of the component for the $i$th test sample. The results are summarized in Table \ref{maretable}. It is seen that the MAREs, for all the 12 components, are below 0.55\%.

\begin{figure}[t!]
	\centering
	\includegraphics[width=0.65\textwidth]{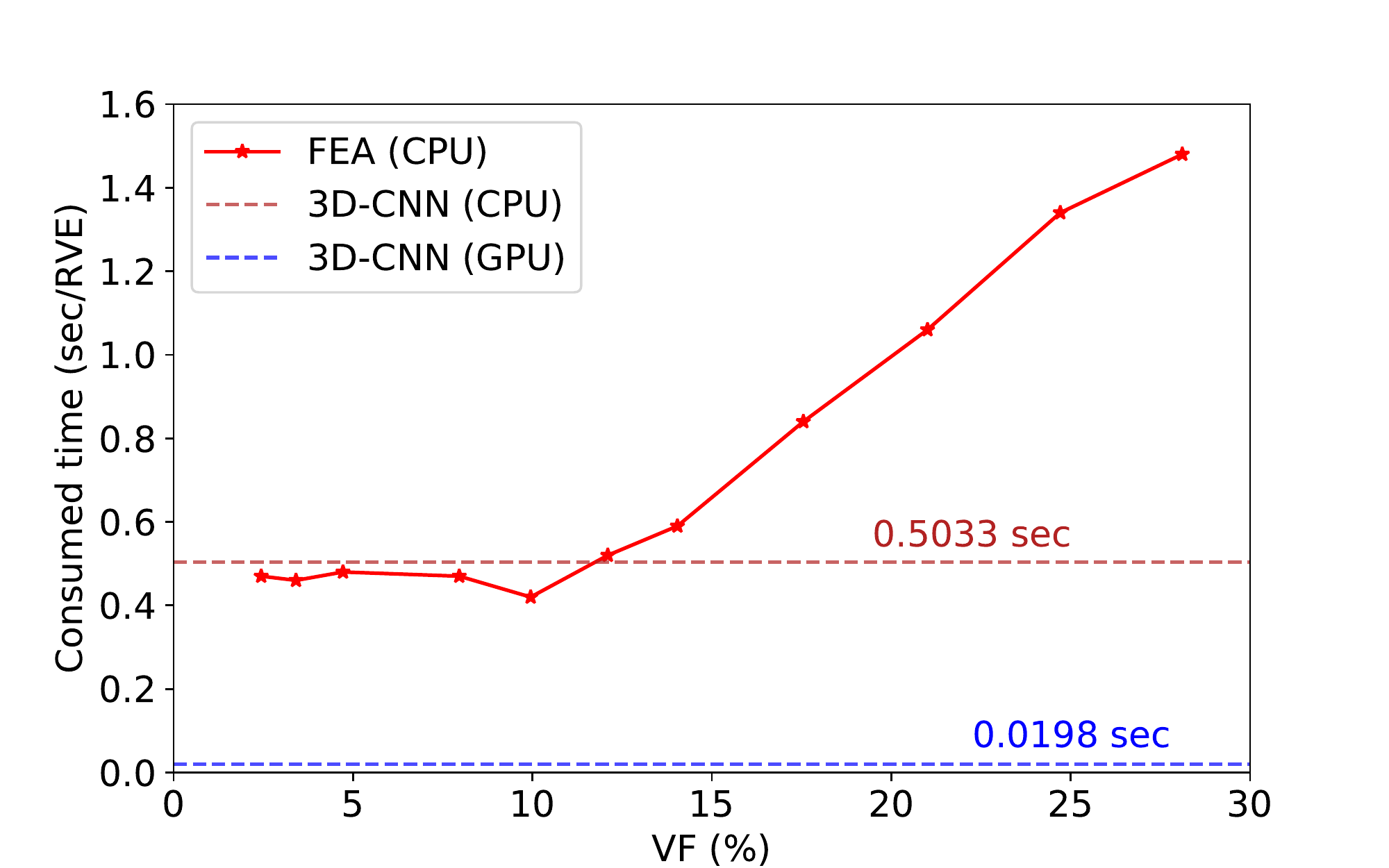}
	\caption{Comparison of the computational time per RVE for 3D-CNN prediction and FEA.}
	\label{timeconsumed}
\end{figure}

\begin{table}[t!]
\caption{MARE on the testing data set.}
\small
\centering
\begin{tabular}{l l l l l l l l l l l l l}
    \hline
    \multicolumn{1}{l}{~} & \multicolumn{1}{l}{\textit{$E_{11}$}} & \multicolumn{1}{l}{\textit{$E_{22}$}} & \multicolumn{1}{l}{\textit{$E_{33}$}} & \multicolumn{1}{l}{\textit{$G_{23}$}} & \multicolumn{1}{l}{\textit{$G_{13}$}} & \multicolumn{1}{l}{\textit{$G_{12}$}} & \multicolumn{1}{l}{$\nu_{21}$} & \multicolumn{1}{l}{$\nu_{31}$} & \multicolumn{1}{l}{$\nu_{12}$} & \multicolumn{1}{l}{$\nu_{32}$} & \multicolumn{1}{l}{$\nu_{13}$} & \multicolumn{1}{l}{$\nu_{23}$}\\
    \hline
    MARE (\%)   & 0.45&0.42&0.47&0.48&0.50&0.53&0.22&0.23&0.24&0.22&0.25&0.22    \\
    \hline
\end{tabular}
\label{maretable}
\end{table}

The efficiency of the proposed 3D-CNN approach is also evaluated by drawing a contrast of computational time between 3D-CNN inference and finite element analysis (FEA), as shown in Fig. \ref{timeconsumed}. Note that the process of inference is defined as the prediction operation on new input data by the trained 3D-CNN model. It is well known that GPU parallelization has been highly exploited on deep learning models in the context of both network training and inference. However, to make the comparison fair, we also collect the averaged CPU time consumed by 3D-CNN by performing inference on the CPU. The configurations of hardware are given in the beginning of Section \ref{results}. It is noted that the CPU time of FEA depends largely on the number of discrete elements of the RVE. In our test, the number of tetrahedral elements in the discretized RVEs increases from 7705 for VF=2.13\% to 26136 for VF=28.22\% to maintain a reliable discretization. We collect the averaged computational time of 10 different RVEs for each fixed VF. For the 3D-CNN inference, however, the computational time is theoretically independent of VF since all the RVEs are sampled with $101\times101\times101$ voxels. We collect the computational time for 300 RVEs with all VF covered. It is seen from Fig. \ref{timeconsumed} that the GPU-based 3D-CNN inference provides 25$\times$ speedup for the low-VF samples and up to 50$\times$ speedup for the highest VF. Even on the CPU, the 3D-CNN beats the traditional FEA for VF greater than 12\%.

Another aspect that cannot be neglected is the computational time for training the 3D-CNN model. For the training dataset with 1400 RVEs considered in this paper, it takes about 35 hours on GPU to achieve a desirable trained model. Nevertheless, it is noticed that the high computational demand for training is one-off which means that, once the model is trained, the inference can be conducted on any upcoming new RVEs that fall into the ensemble. Even if the new RVE comes from another type of composite, the transferability of the trained 3D-CNN, discussed in Section \ref{transferlearning}, will largely reduce the time expense. We will verify that transfer learning makes the 3D-CNN extremely convenient for adding supplementary data or training a model for new datasets to account for new scenarios and enhance the generalizability of the trained model.

\begin{figure}[t!]
\centering 
\includegraphics[width=1.0\textwidth]{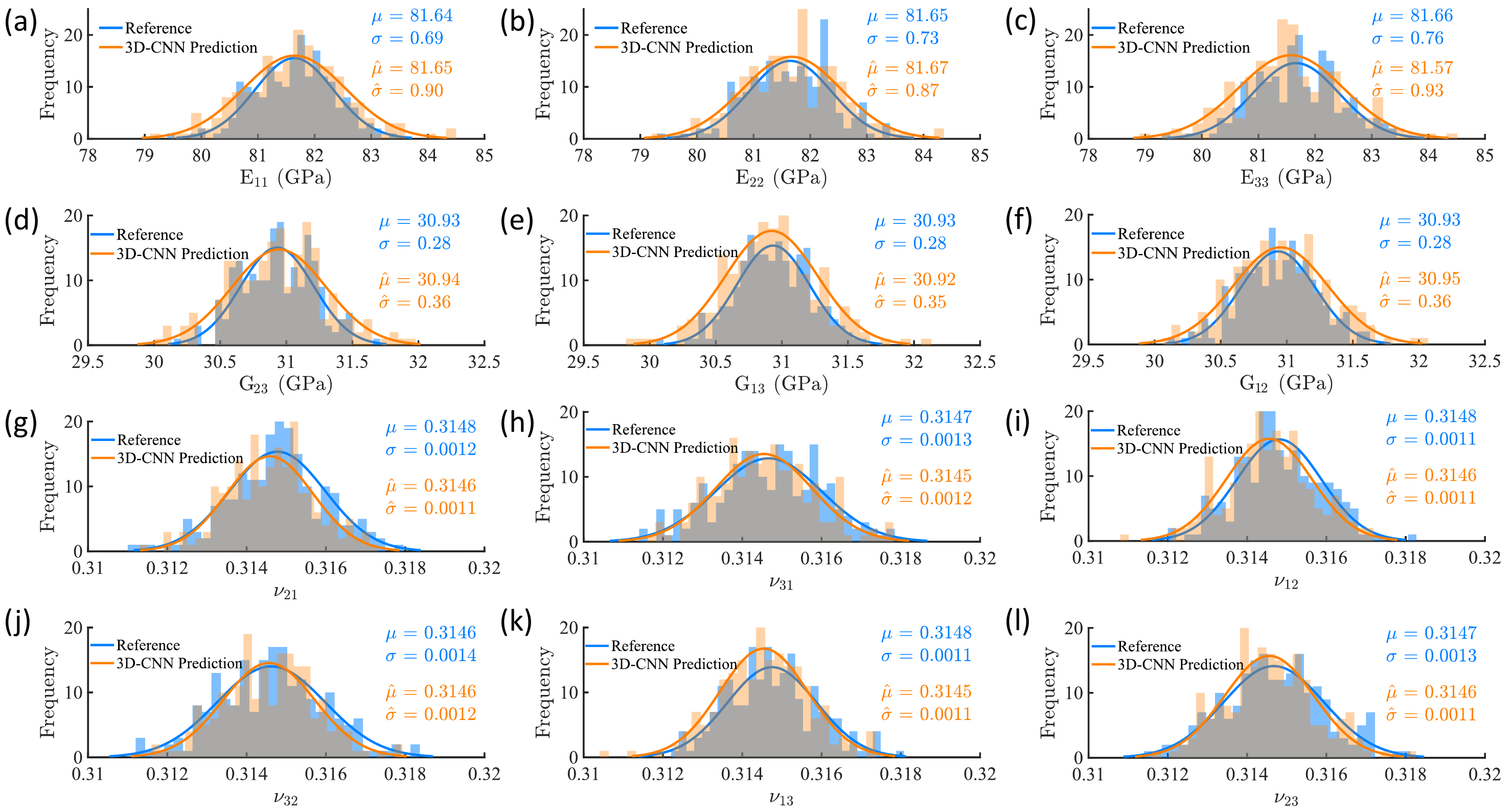}
\caption{Distribution of effective properties of the 3D-CNN prediction and FEA result for the dataset of VF=7\%: (a) $E_{11}$  (b) $E_{22}$ (c) $E_{33}$ (d) $G_{23}$ (e) $G_{13}$ (f) $G_{12}$ (g) $\nu_{21}$ (h) $\nu_{31}$ (i) $\nu_{12}$ (j) $\nu_{32}$ (k) $\nu_{13}$ (l) $\nu_{23}$. }
\label{comparison}
\end{figure}

\subsection{Uncertainty quantification}\label{UQ}
Modelling of natural composites is usually characterized with uncertainty. The uncertainty may come from the measurement error, microstructural randomness, mixture of materials and some other natural (or artificial) systems. Predicting the effective properties in a probabilistic/statistical sense, such as obtaining the mean value and standard deviation (SD), would provide a better reference for engineering and designing materials. 

Strictly speaking, the output of a trained 3D-CNN is deterministic for a given input. Therefore, the uncertainty of the 3D-CNN output is largely affected by the variance of the input. To verify that our 3D-CNN model is capable of preserving the uncertainty of the effective properties for the particle reinforced composite, we manually introduce the uncertainty into the dataset to be evaluated in the framework of Monte Carlo simulation. In particular, we generate a group of RVE samples VF following Gaussian distributions (e.g., mean of $7\%$, $14\%$ and $21\%$ for three configurations, and identical standard deviation of 0.7\%). In each configuration, 200 RVEs are generated. The details for the uncertainty quantification (UQ) dataset are listed in Table \ref{uq_para}.

\begin{table}[t!]
\caption{VF parameters used for uncertainty quantification.}
\small
\centering
\begin{tabular}{l l l l l}
    \hline
    \multicolumn{1}{l}{Mean ($\mu$, \%)} & \multicolumn{1}{l}{SD ($\sigma$, \%)} & \multicolumn{1}{l}{Number of RVE Samples}\\ 
    \hline
    7   & 0.7  &   200    \\
    14   & 0.7  &  200    \\
    21   & 0.7  &   200    \\
    \hline
\end{tabular}
\label{uq_para}
\end{table}

Fig. \ref{comparison} presents the predicted distributions of the modulus and Poisson's ratio components in comparison with the reference ground truth. These histograms are fitted by Gaussian distributions whose mean and standard deviation parameters are also listed. It can be seen that the trained 3D-CNN produces very satisfactory prediction of the probabilistic distributions, e.g., the errors for the mean value of all the components are less than 1\% while the predicted standard deviations are also very close, but slightly larger than, the ground truth values.

\begin{figure}[t!]
\centering 
\includegraphics[width=1.0\textwidth]{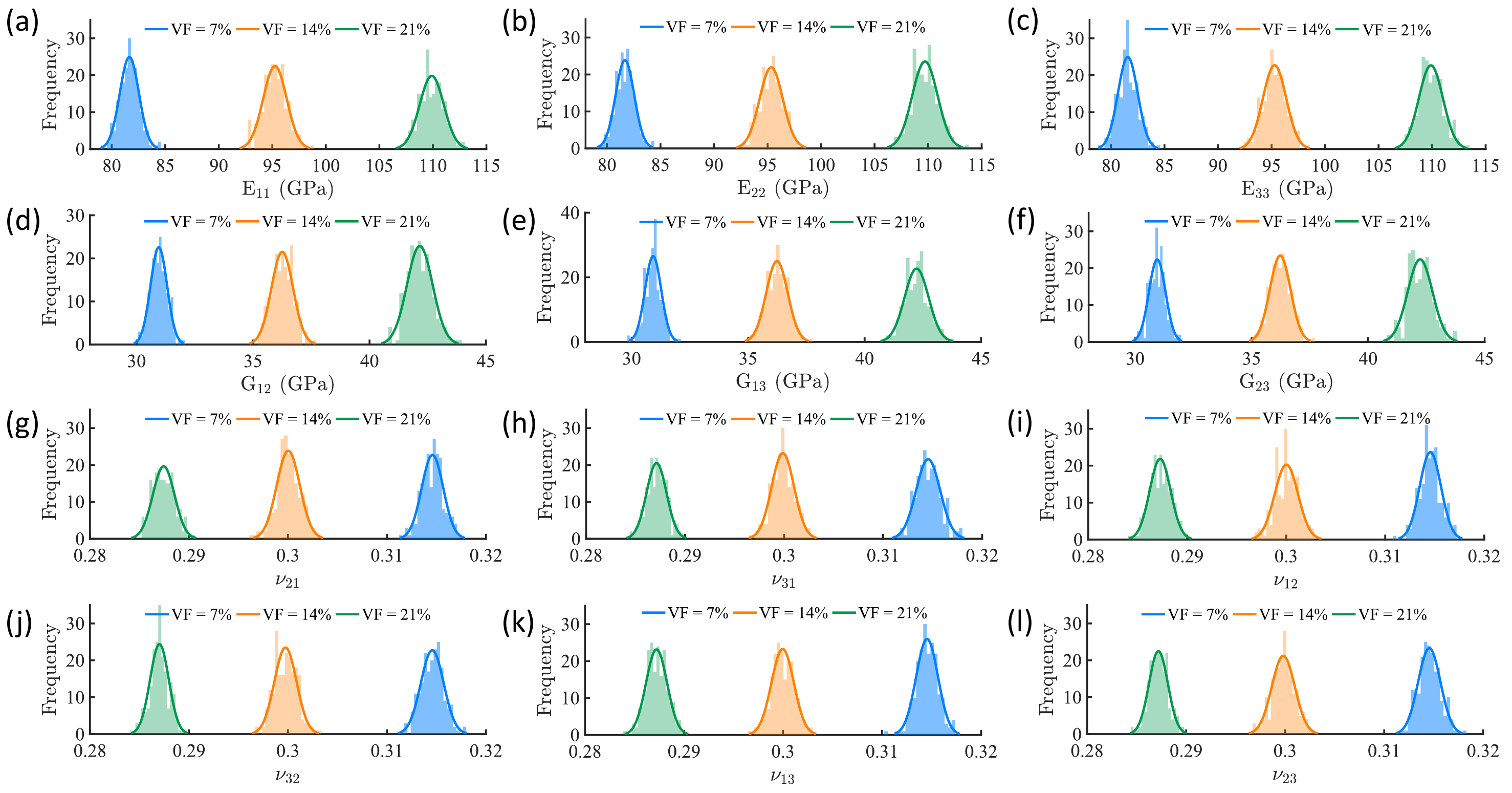}
\caption{Distribution of predicted effective properties for RVEs with mean VF of 7\%, 14\% and 28\%: (a) $E_{11}$  (b) $E_{22}$ (c) $E_{33}$ (d) $G_{23}$ (e) $G_{13}$ (f) $G_{12}$ (g) $\nu_{21}$ (h) $\nu_{31}$ (i) $\nu_{12}$ (j) $\nu_{32}$ (k) $\nu_{13}$ (l) $\nu_{23}$.}
\label{allprediction}
\end{figure}

The predicted distributions of effective properties for three VF cases are shown in Fig. \ref{allprediction}. It is obvious that the modulus components are positively correlated to the VF while the Poisson's ratio components are on the contrary, which are in accordance with the Voigt/Reuss models \cite{voigt, reuss1929}. In a word, the 3D-CNN's ability to reproduce the probabilistic distribution of the effective properties, together with its high computational efficiency (as discussed in Section \ref{discussion3D-CNN}), will make it a promising approach for probabilistic design of engineering composites \cite{du2002efficient, chen2006probabilistic}.

\begin{figure}[t!]
\centering
\subfigure[VF = 6.26\%]{
\begin{minipage}[t]{0.32\linewidth}
\centering
\includegraphics[width=\linewidth]{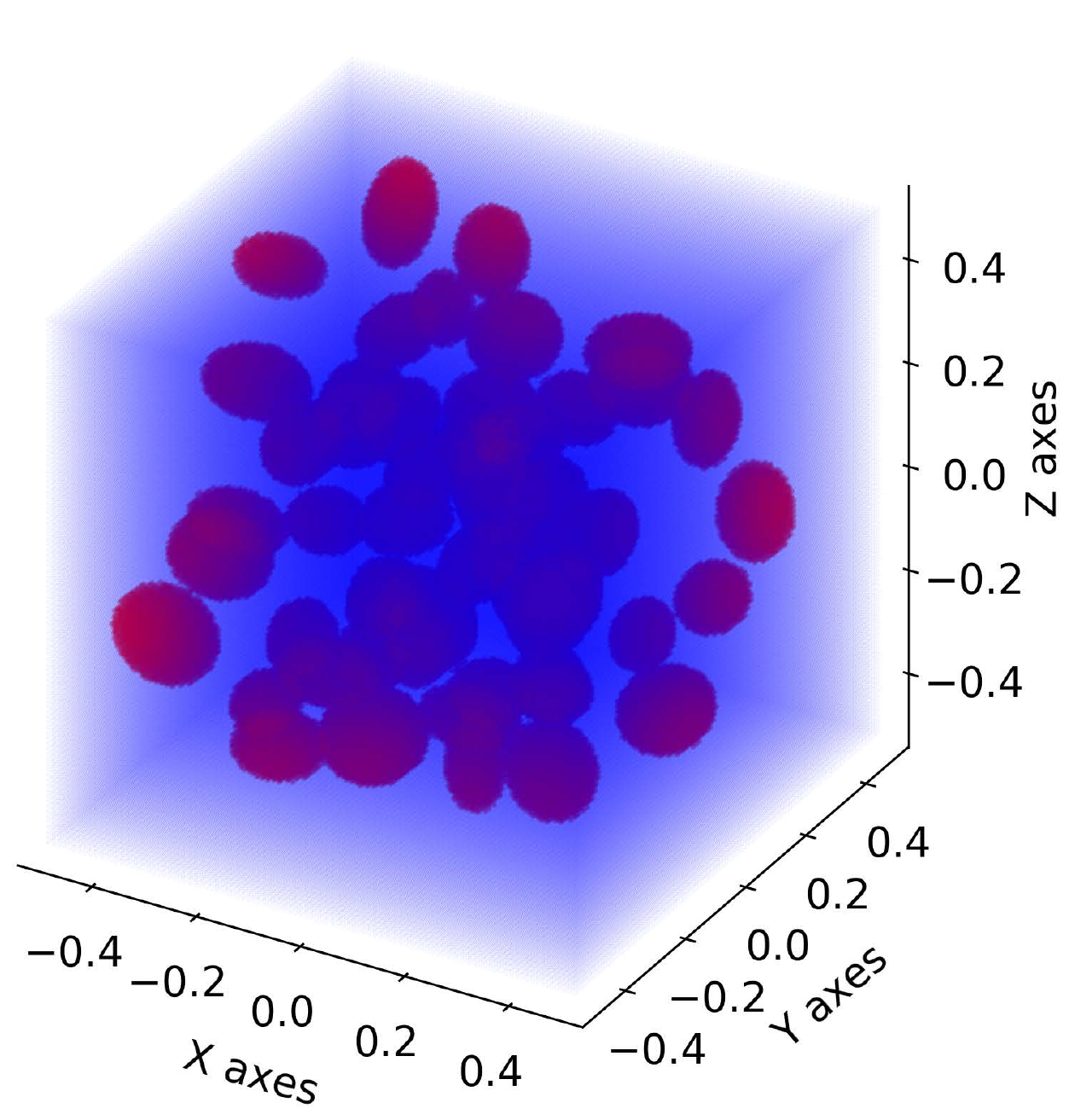}
\end{minipage}%
}%
\subfigure[VF = 14.67\%]{
\begin{minipage}[t]{0.32\linewidth}
\centering
\includegraphics[width=\linewidth]{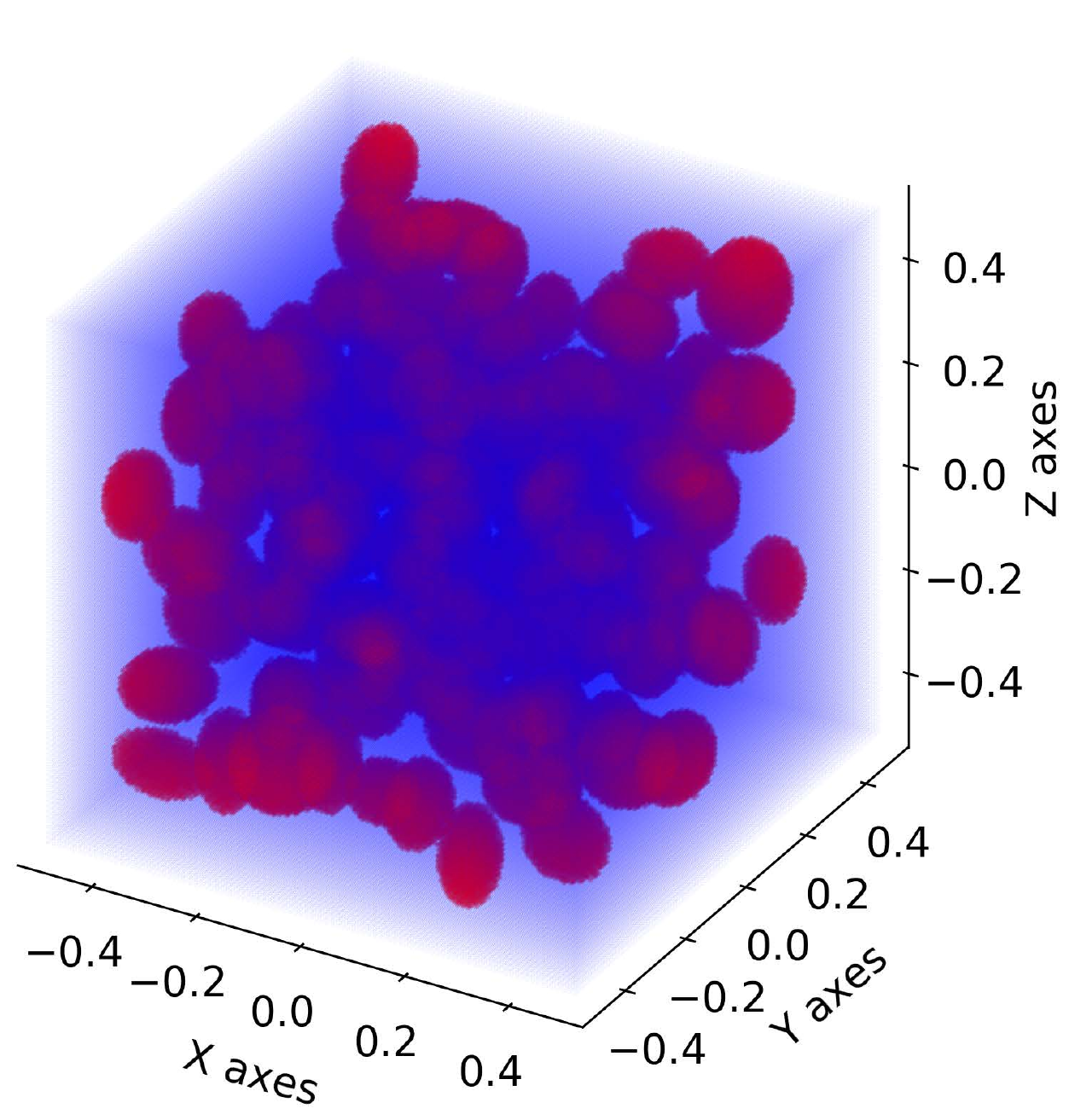}
\end{minipage}%
}%
\hfill
\centering
\subfigure[VF = 20.79\%]{
\begin{minipage}[t]{0.32\linewidth}
\centering
\includegraphics[width=\linewidth]{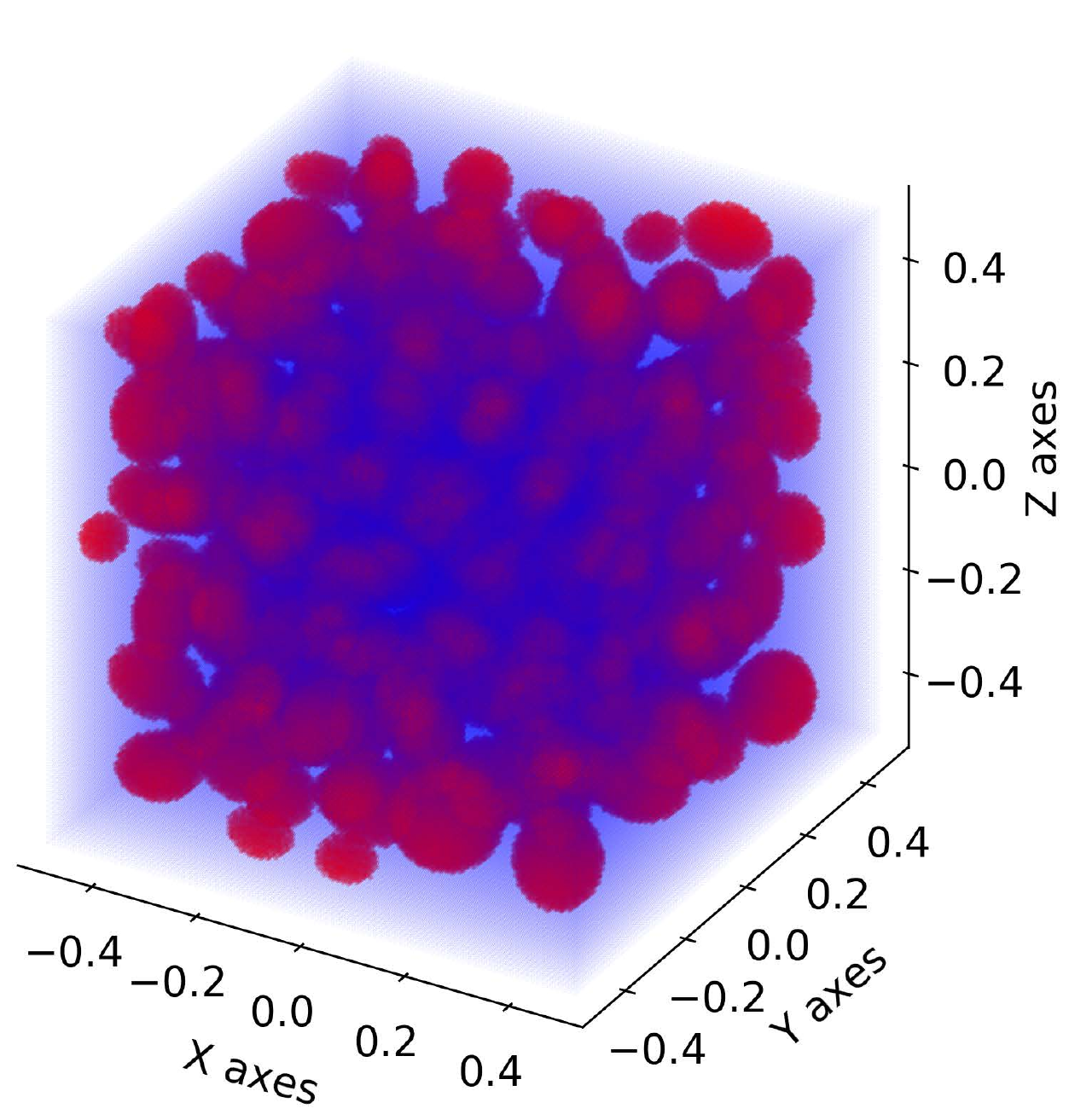}
\end{minipage}
}%
\caption{Sampled phase voxel for RVEs with ellipsoidal inclusions}\label{elliprve}
\end{figure}

\subsection{Transferability of the trained model} \label{transferlearning}
A major assumption required by lots of DL approaches is that the training data and future data must be from the same generator or source. In other words, they must be in the same feature space and follow the same distribution \cite{pan2009survey}. In many real-world applications, this assumption may not hold. In these cases, if the knowledge learned by the DL model can be transferred, it will largely reduce the effort on retraining the model on new datasets. The transferability refers to the convenience of transferring the learned knowledge from a trained model to a different but related problem. Transfer learning is usually achieved through transfer the pre-trained model to a new model with additional trainable parameters relying on new datasets of interest (e.g., adding additional layers to the trained network while fixing the transferred network parameters from the original model). The need for transferring learning arises when the acquired data can be easily outdated or when the target data is intractable (or costly) to obtain but a less rich dataset is available.

\begin{figure}[t!]
	\centering
	\includegraphics[width=10cm]{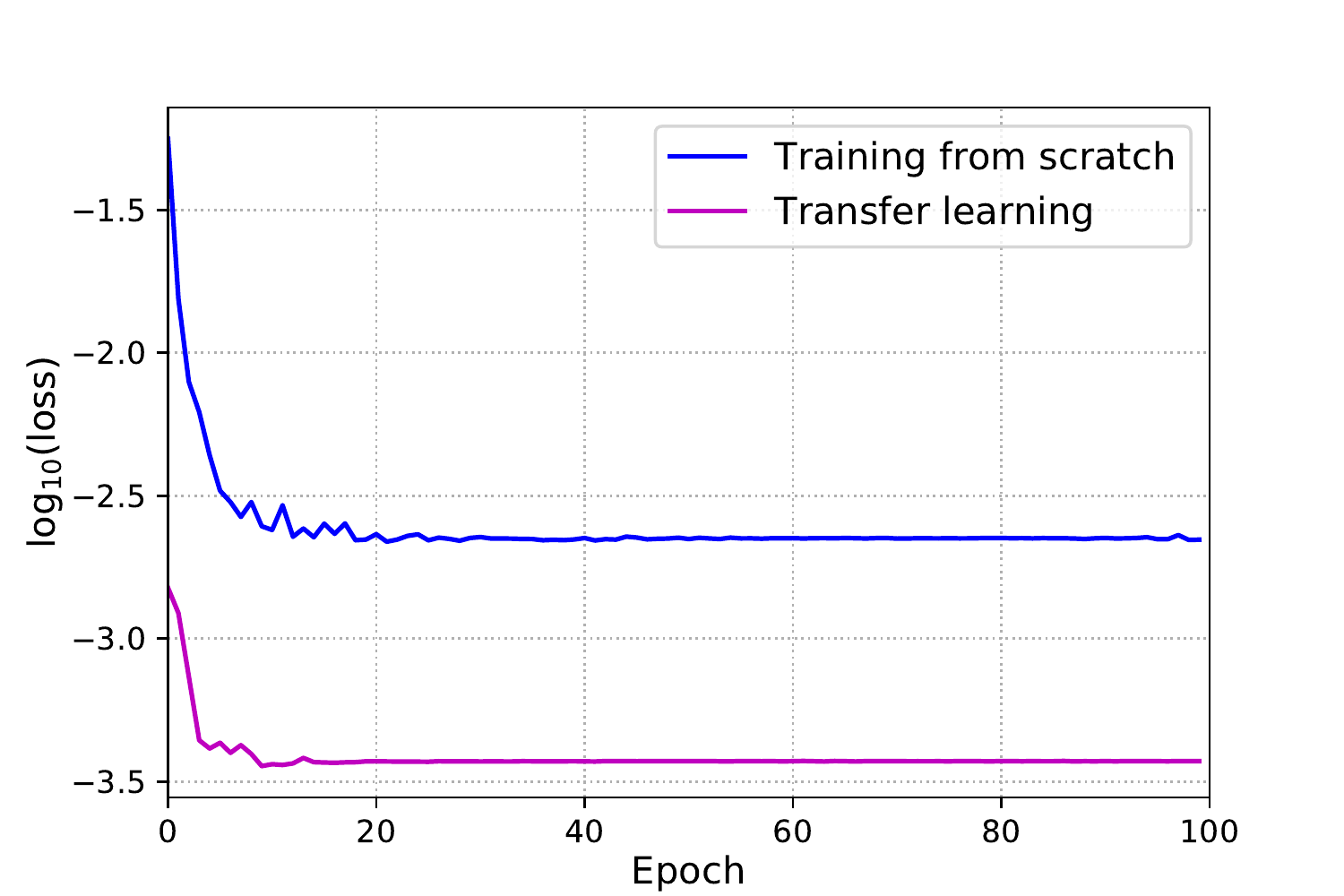}
	\caption{Comparison of the learning curve}
	\label{learningcurve}
\end{figure}

\begin{figure}[t!]
	\centering
	\includegraphics[width=16cm]{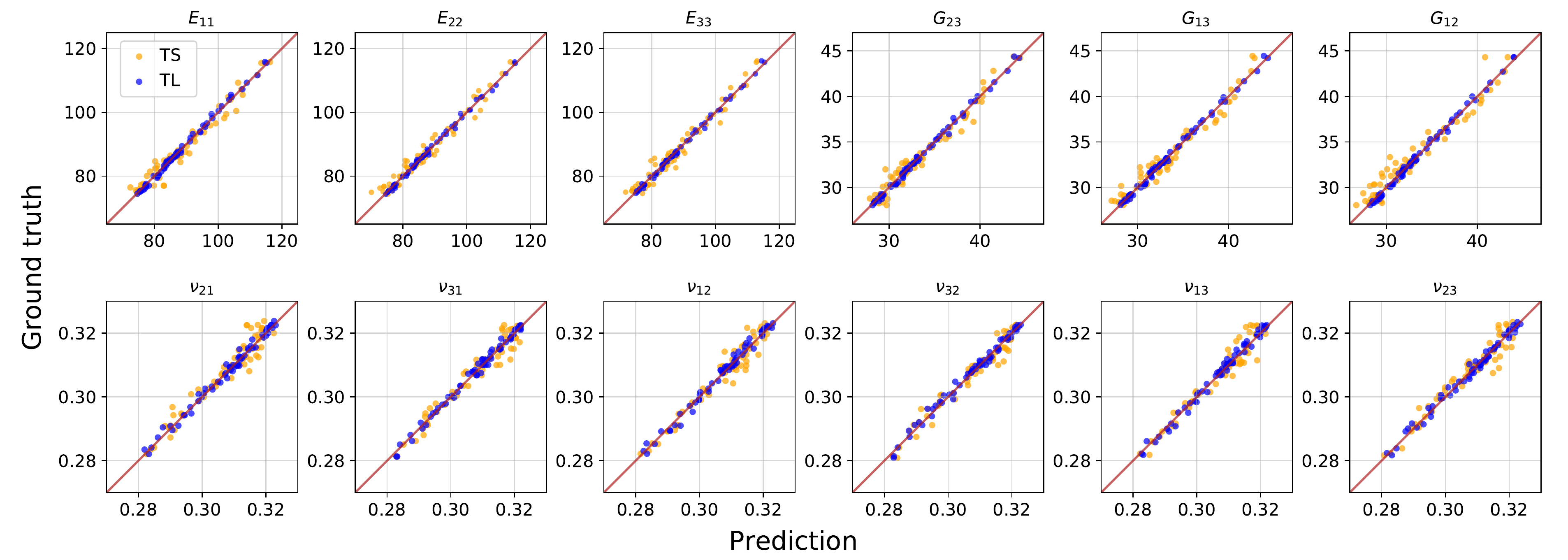}
	\caption{Comparison of the prediction performance for trained from scratch (TS) model and transfer learning (TL) model.}
	\label{transferlearningperfor}
\end{figure}

To examine the transferability of the previously trained 3D-CNN model, we consider a new dataset of RVEs with ellipsoidal inclusions. The major and minor radius of the ellipsoids are randomly generated within the interval $[0.05,0.1]$ independently. The overall range of the VF is the same as the previous data set (e.g., 2\%-28\%). Following the similar manner in Section \ref{pre}, a much smaller dataset with only 320 samples is generated with the sample number as an exponential function of the VF. The entire data is divided into training, validation and testing set with the ratio of 200:60:60. We transfer the trained 3D-CNN model with the architecture described in Case 2 as shown in Table \ref{performance}, and establish a new 3D-CNN network by adding one additional convolution layer before flattening, e.g., Conv(32, 5), and activate the trainable parameters in the last FC layer (see Fig. \ref{CNN_ARCH}). We try to generalize the trained 3D-CNN for RVEs with spherical inclusions to the case of ellipsoidal inclusions (see Fig. \ref{elliprve} for example). The transfer learning (TL) model fine tuned with new dataset is compared with the model trained from scratch (TS) with regard to the learning curve and prediction performance. 

The learning curves for both cases (e.g., TL \emph{vs.} TS) are shown in Fig. \ref{learningcurve} where the $x$-axis denotes the epoch and $y$-axis denotes the loss function value. It can be seen that the initial loss is much lower for the TL model which indicates that the transferred model for sphere inclusions can already well capture the latent features for RVEs with ellipsoidal inclusions. The asymptote for the TS convergence curve is much higher than that of the TL model. Given a small amount of training dataset, the TL model converges much faster, e.g., only taking dozens of epochs for the loss to decrease to $3.7\times10^{-4}$ which is close to our best model ($2.79\times10^{-4}$) discussed in Section \ref{parameterictest}. It demonstrates that we can successfully transfer the knowledge of as well as fine tune a pre-trained 3D-CNN model to achieve a good accuracy at a particular low training expense. Therefore, the transfer learning might help overcome problems such as lack of the data and high computational cost for training a large size model. These challenges are critical especially in field measurements where rich RVEs data are costly to obtain. The prediction performance of these both TL and TS models are compared in Fig. \ref{transferlearningperfor}. It is evident that the TL model outperforms the TS model no matter in the bias or variance of the effective properties. The averaged MARE for the TL and TS models on all the components are 0.43\% and 1.36\%, respectively.

\section{Conclusions}\label{conc}

In this paper, a 3D-CNN approach is proposed for determining the effective/homogenized properties of heterogeneous materials. In particular, we consider RVEs reinforced by reandomly distributed particle inclusion (e.g., spherical and elliptical inclusions). The geometries of the RVEs are generated using the Hierarchical Random Sequential Adsorption (HRSA) algorithm \cite{bai2014auto} and labeled for training the 3D-CNN model via FEA-based linear homogenization. The proposed 3D-CNN architecture consists of multiple hidden 3D convolution layers, pooling operation, flattening and FC layers. A parametric study of the network hyperparameters has been conducted to determine optimal network architecture with the best inference performance. The proposed approach was tested on a series of numerical experiments in the context of inference accuracy, computational efficiency, uncertainty quantification (UQ) ability and transferability. Results show promising potential of the proposed approach to advance efficient design and analysis of heterogeneous composite materials composed of representative microstructures. 

It is worth mentioning that the comparison with the FEA results shows that the 3D-CNN model can reproduce the effective material properties with a high accuracy (e.g., the maximum prediction error around 0.5\%). Also, the 3D-CNN demonstrates advantages regarding the computational efficiency for the model inference over the traditional FEA, which could achieve a speed-up from 25$\times$ to 50$\times$ on GPU operation. In addition, the UQ study verifies the trained 3D-CNN is capable of accurately predicting probabilistic distributions of the effective material properties, in the framework of Monte Carlo simulation, when uncertain inputs are provided. 

In summary, the proposed 3D-CNN is characterized with the following benefits: (1) It provides an end-to-end solution for predicting the effective material properties from 3D phase voxels which can be obtained via parametric modeling, advanced imaging techniques such as X-ray micro-topography and 3D atom probe; (2) It is able to reproduce the effective properties with a high accuracy and computational efficiency, which would empower a faster product design iteration or design optimization for composite materials; (3) The 3D-CNN model preserves the probabilistic distribution of effective material properties for the input with uncertainty. This feature makes the 3D-CNN a promising approach for probabilistic engineering design; (4) The knowledge learned by the 3D-CNN model can be easily transferred to a different type of composite at a very low training expense, in which a good prediction performance can still be achieved even on a new dataset of small size with the help of transfer learning. This particular characteristic becomes significant when RVEs data are costly to obtain.

Nevertheless, there remain some issues of interest on the 3D-CNN model to be studied in the future, that include, for example: (1) investigating the universality of transfer learning on other heterogeneous materials such as fiber-reinforced or polymer composites; (2) extending the current 3D-CNN to model composites with nonlinear material properties (to this end, the load condition on each RVE must be considered as part of the input for the networks); (3) applying the trained model or retraining a generative model for  microstructure generation with desired effective properties \cite{yang18gan, li2018GAN}. 

\section*{Acknowledgement}
The authors would like to thank Dr. Hao Sun and Dr. Ruiyang Zhang, from the Department of Civil and Environmental Engineering at Northeastern University, for their constructive suggestions and comments on designing the proposed network.

\section*{Data Availability}
The datasets and computer codes are available upon request from the authors.

\section*{References}
\bibliographystyle{elsarticle-num}
\bibliography{refs}

\begin{thebibliography}{10}
\expandafter\ifx\csname url\endcsname\relax
  \def\url#1{\texttt{#1}}\fi
\expandafter\ifx\csname urlprefix\endcsname\relax\def\urlprefix{URL }\fi
\expandafter\ifx\csname href\endcsname\relax
  \def\href#1#2{#2} \def\path#1{#1}\fi

\bibitem{hornung2012homogenization}
U.~Hornung, Homogenization and porous media, Vol.~6, Springer Science \&
  Business Media, 2012.

\bibitem{aboudi2012}
J.~Aboudi, S.~M. Arnold, B.~A. Bednarcyk, Micromechanics of composite
  materials: a generalized multiscale analysis approach, Butterworth-Heinemann,
  2012.

\bibitem{voigt}
W.~Voigt, Ueber die beziehung zwischen den beiden elasticit{\"a}tsconstanten
  isotroper k{\"o}rper, Annalen der Physik 274~(12) (1889) 573--587.

\bibitem{reuss1929}
A.~Reu{\ss}, Berechnung der flie{\ss}grenze von mischkristallen auf grund der
  plastizit{\"a}tsbedingung f{\"u}r einkristalle., ZAMM-Journal of Applied
  Mathematics and Mechanics/Zeitschrift f{\"u}r Angewandte Mathematik und
  Mechanik 9~(1) (1929) 49--58.

\bibitem{aboudi2004gmc}
J.~Aboudi, The generalized method of cells and high-fidelity generalized method
  of cells micromechanical models—a review, Mechanics of Advanced Materials
  and Structures 11~(4-5) (2004) 329--366.

\bibitem{scs1968}
Z.~Hashin, Assessment of the self consistent scheme approximation: conductivity
  of particulate composites, Journal of Composite Materials 2~(3) (1968)
  284--300.

\bibitem{scs1978}
M.~Berveiller, A.~Zaoui, An extension of the self-consistent scheme to
  plastically-flowing polycrystals, Journal of the Mechanics and Physics of
  Solids 26~(5-6) (1978) 325--344.

\bibitem{mori1973}
T.~Mori, K.~Tanaka, Average stress in matrix and average elastic energy of
  materials with misfitting inclusions, Acta Metallurgica 21~(5) (1973)
  571--574.

\bibitem{feyel1999fe2}
F.~Feyel, Multiscale {$\text{FE}^2$} elastoviscoplastic analysis of composite
  structures, Computational Materials Science 16~(1-4) (1999) 344--354.

\bibitem{feyel2000fe2}
F.~Feyel, J.-L. Chaboche, {$\text{FE}^2$} multiscale approach for modelling the
  elastoviscoplastic behaviour of long fibre {SiC/Ti} composite materials,
  Computer Methods in Applied Mechanics and Engineering 183~(3-4) (2000)
  309--330.

\bibitem{feyel2003fe2}
F.~Feyel, A multilevel finite element method ({$\text{FE}^2$}) to describe the
  response of highly non-linear structures using generalized continua, Computer
  Methods in Applied Mechanics and Engineering 192~(28-30) (2003) 3233--3244.

\bibitem{miehe2002fe}
C.~Miehe, J.~Schotte, M.~Lambrecht, Homogenization of inelastic solid materials
  at finite strains based on incremental minimization principles. application
  to the texture analysis of polycrystals, Journal of the Mechanics and Physics
  of Solids 50~(10) (2002) 2123--2167.

\bibitem{smit1998fe}
R.~Smit, W.~Brekelmans, H.~Meijer, Prediction of the mechanical behavior of
  nonlinear heterogeneous systems by multi-level finite element modeling,
  Computer Methods in Applied Mechanics and Engineering 155~(1-2) (1998)
  181--192.

\bibitem{terada2001fe}
K.~Terada, N.~Kikuchi, A class of general algorithms for multi-scale analyses
  of heterogeneous media, Computer Methods in Applied Mechanics and Engineering
  190~(40-41) (2001) 5427--5464.

\bibitem{kaminski1999bem}
M.~Kami{\'n}ski, Boundary element method homogenization of the periodic linear
  elastic fiber composites, Engineering Analysis with Boundary Elements 23~(10)
  (1999) 815--823.

\bibitem{okada2001bem}
H.~Okada, Y.~Fukui, N.~Kumazawa, Homogenization method for heterogeneous
  material based on boundary element method, Computers \& Structures 79~(20-21)
  (2001) 1987--2007.

\bibitem{lee2011fft}
S.-B. Lee, R.~Lebensohn, A.~D. Rollett, Modeling the viscoplastic
  micromechanical response of two-phase materials using fast fourier
  transforms, International Journal of Plasticity 27~(5) (2011) 707--727.

\bibitem{eisenlohr2013fft}
P.~Eisenlohr, M.~Diehl, R.~A. Lebensohn, F.~Roters, A spectral method solution
  to crystal elasto-viscoplasticity at finite strains, International Journal of
  Plasticity 46 (2013) 37--53.

\bibitem{kanoute2009review}
P.~Kanout{\'e}, D.~Boso, J.~Chaboche, B.~Schrefler, Multiscale methods for
  composites: a review, Archives of Computational Methods in Engineering 16~(1)
  (2009) 31--75.

\bibitem{yuan2008homo}
Z.~Yuan, J.~Fish, Toward realization of computational homogenization in
  practice, International Journal for Numerical Methods in Engineering 73~(3)
  (2008) 361--380.

\bibitem{hain2008numerical}
M.~Hain, P.~Wriggers, Numerical homogenization of hardened cement paste,
  Computational Mechanics 42~(2) (2008) 197--212.

\bibitem{liu2014regularized}
Y.~Liu, V.~Filonova, N.~Hu, Z.~Yuan, J.~Fish, Z.~Yuan, T.~Belytschko, A
  regularized phenomenological multiscale damage model, International Journal
  for Numerical Methods in Engineering 99~(12) (2014) 867--887.

\bibitem{liu2016nonlocal}
Y.~Liu, W.~Sun, Z.~Yuan, J.~Fish, A nonlocal multiscale discrete-continuum
  model for predicting mechanical behavior of granular materials, International
  Journal for Numerical Methods in Engineering 106~(2) (2016) 129--160.

\bibitem{fritzen2018two}
F.~Fritzen, O.~Kunc, Two-stage data-driven homogenization for nonlinear solids
  using a reduced order model, European Journal of Mechanics-A/Solids 69 (2018)
  201--220.

\bibitem{olson1997computational}
G.~B. Olson, Computational design of hierarchically structured materials,
  Science 277~(5330) (1997) 1237--1242.

\bibitem{lookman2019active}
T.~Lookman, P.~V. Balachandran, D.~Xue, R.~Yuan, Active learning in materials
  science with emphasis on adaptive sampling using uncertainties for targeted
  design, npj Computational Materials 5~(1) (2019) 21.

\bibitem{fujii2001composite}
D.~Fujii, B.~Chen, N.~Kikuchi, Composite material design of two-dimensional
  structures using the homogenization design method, International Journal for
  Numerical Methods in Engineering 50~(9) (2001) 2031--2051.

\bibitem{landi2010mks}
G.~Landi, S.~R. Niezgoda, S.~R. Kalidindi, Multi-scale modeling of elastic
  response of three-dimensional voxel-based microstructure datasets using novel
  dft-based knowledge systems, Acta Materialia 58~(7) (2010) 2716--2725.

\bibitem{fast2011mks}
T.~Fast, S.~R. Niezgoda, S.~R. Kalidindi, A new framework for computationally
  efficient structure--structure evolution linkages to facilitate high-fidelity
  scale bridging in multi-scale materials models, Acta Materialia 59~(2) (2011)
  699--707.

\bibitem{kalidindi2015mks}
S.~R. Kalidindi, Hierarchical materials informatics: novel analytics for
  materials data, Elsevier, 2015.

\bibitem{kroner1986statistical}
E.~Kr{\"o}ner, Statistical modelling, in: Modelling small deformations of
  polycrystals, Springer, 1986, pp. 229--291.

\bibitem{yabansu2014mks}
Y.~C. Yabansu, D.~K. Patel, S.~R. Kalidindi, Calibrated localization
  relationships for elastic response of polycrystalline aggregates, Acta
  Materialia 81 (2014) 151--160.

\bibitem{gupta2015mks}
A.~Gupta, A.~Cecen, S.~Goyal, A.~K. Singh, S.~R. Kalidindi, Structure--property
  linkages using a data science approach: application to a non-metallic
  inclusion/steel composite system, Acta Materialia 91 (2015) 239--254.

\bibitem{lecun2015deep}
Y.~LeCun, Y.~Bengio, G.~Hinton, Deep learning, Nature 521~(7553) (2015) 436.

\bibitem{zheng2016}
H.~Zheng, L.~Fang, M.~Ji, M.~Strese, Y.~{\"O}zer, E.~Steinbach, Deep learning
  for surface material classification using haptic and visual information, IEEE
  Transactions on Multimedia 18~(12) (2016) 2407--2416.

\bibitem{Bell2015}
S.~Bell, P.~Upchurch, N.~Snavely, K.~Bala, Material recognition in the wild
  with the materials in context database, in: The IEEE Conference on Computer
  Vision and Pattern Recognition (CVPR), 2015.

\bibitem{masci2012}
J.~Masci, U.~Meier, D.~Ciresan, J.~Schmidhuber, G.~Fricout, Steel defect
  classification with max-pooling convolutional neural networks, in: The 2012
  International Joint Conference on Neural Networks (IJCNN), IEEE, 2012, pp.
  1--6.

\bibitem{cha2017deep}
Y.-J. Cha, W.~Choi, O.~B{\"u}y{\"u}k{\"o}zt{\"u}rk, Deep learning-based crack
  damage detection using convolutional neural networks, Computer-Aided Civil
  and Infrastructure Engineering 32~(5) (2017) 361--378.

\bibitem{faghih2016deep}
S.~Faghih-Roohi, S.~Hajizadeh, A.~N{\'u}{\~n}ez, R.~Babuska, B.~De~Schutter,
  Deep convolutional neural networks for detection of rail surface defects, in:
  2016 International joint conference on neural networks (IJCNN), IEEE, 2016,
  pp. 2584--2589.

\bibitem{azimi2018}
S.~M. Azimi, D.~Britz, M.~Engstler, M.~Fritz, F.~M{\"u}cklich, Advanced steel
  microstructural classification by deep learning methods, Scientific Reports
  8~(1) (2018) 2128.

\bibitem{chowdhury2016image}
A.~Chowdhury, E.~Kautz, B.~Yener, D.~Lewis, Image driven machine learning
  methods for microstructure recognition, Computational Materials Science 123
  (2016) 176--187.

\bibitem{li2018transfer}
X.~Li, Y.~Zhang, H.~Zhao, C.~Burkhart, L.~C. Brinson, W.~Chen, A transfer
  learning approach for microstructure reconstruction and structure-property
  predictions, Scientific Reports 8.

\bibitem{li2018GAN}
X.~Li, Z.~Yang, L.~C. Brinson, A.~Choudhary, A.~Agrawal, W.~Chen, A deep
  adversarial learning methodology for designing microstructural material
  systems, in: ASME 2018 international design engineering technical conferences
  and computers and information in engineering conference, American Society of
  Mechanical Engineers, 2018, pp. V02BT03A008--V02BT03A008.

\bibitem{yeh1998modeling}
I.-C. Yeh, Modeling of strength of high-performance concrete using artificial
  neural networks, Cement and Concrete Research 28~(12) (1998) 1797--1808.

\bibitem{lu2018data}
X.~Lu, D.~G. Giovanis, J.~Yvonnet, V.~Papadopoulos, F.~Detrez, J.~Bai, A
  data-driven computational homogenization method based on neural networks for
  the nonlinear anisotropic electrical response of graphene/polymer
  nanocomposites, Computational Mechanics (2018) 1--15.

\bibitem{le2015computational}
B.~Le, J.~Yvonnet, Q.-C. He, Computational homogenization of nonlinear elastic
  materials using neural networks, International Journal for Numerical Methods
  in Engineering 104~(12) (2015) 1061--1084.

\bibitem{bhattacharjee2016nonlinear}
S.~Bhattacharjee, K.~Matou{\v{s}}, A nonlinear manifold-based reduced order
  model for multiscale analysis of heterogeneous hyperelastic materials,
  Journal of Computational Physics 313 (2016) 635--653.

\bibitem{yang18gan}
Z.~Yang, X.~Li, L.~C. Brinson, A.~N. Choudhary, W.~Chen, A.~Agrawal,
  Microstructural materials design via deep adversarial learning methodology,
  Journal of Mechanical Design 140~(11) (2018) 111416.

\bibitem{cang2017microstructure}
R.~Cang, Y.~Xu, S.~Chen, Y.~Liu, Y.~Jiao, M.~Yi~Ren, Microstructure
  representation and reconstruction of heterogeneous materials via deep belief
  network for computational material design, Journal of Mechanical Design
  139~(7).

\bibitem{bost2016}
R.~Bostanabad, A.~T. Bui, W.~Xie, D.~W. Apley, W.~Chen, Stochastic
  microstructure characterization and reconstruction via supervised learning,
  Acta Materialia 103 (2016) 89--102.

\bibitem{yang2018dl}
Z.~Yang, Y.~C. Yabansu, R.~Al-Bahrani, W.-k. Liao, A.~N. Choudhary, S.~R.
  Kalidindi, A.~Agrawal, Deep learning approaches for mining structure-property
  linkages in high contrast composites from simulation datasets, Computational
  Materials Science 151 (2018) 278--287.

\bibitem{ji3DCNN}
S.~Ji, W.~Xu, M.~Yang, K.~Yu, 3d convolutional neural networks for human action
  recognition, IEEE Transactions on Pattern Analysis and Machine Intelligence
  35~(1) (2012) 221--231.

\bibitem{maturana3dCNN}
D.~Maturana, S.~Scherer, Voxnet: A {3D} convolutional neural network for
  real-time object recognition, in: 2015 IEEE/RSJ International Conference on
  Intelligent Robots and Systems (IROS), IEEE, 2015, pp. 922--928.

\bibitem{kamnitsas3dcnn}
K.~Kamnitsas, C.~Ledig, V.~F. Newcombe, J.~P. Simpson, A.~D. Kane, D.~K. Menon,
  D.~Rueckert, B.~Glocker, Efficient multi-scale {3D CNN} with fully connected
  crf for accurate brain lesion segmentation, Medical Image Analysis 36 (2017)
  61--78.

\bibitem{bai2014auto}
M.~Bailakanavar, Y.~Liu, J.~Fish, Y.~Zheng, Automated modeling of random
  inclusion composites, Engineering with Computers 30~(4) (2014) 609--625.

\bibitem{guedes1990homo}
J.~Guedes, N.~Kikuchi, Preprocessing and postprocessing for materials based on
  the homogenization method with adaptive finite element methods, Computer
  Methods in Applied Mechanics and Engineering 83~(2) (1990) 143--198.

\bibitem{ripley1996pattern}
B.~D. Ripley, N.~Hjort, Pattern recognition and neural networks, Cambridge
  University Press, 1996.

\bibitem{stienon2009xray}
A.~Stienon, A.~Fazekas, J.-Y. Buffiere, A.~Vincent, P.~Daguier, F.~Merchi, A
  new methodology based on {X-ray} micro-tomography to estimate stress
  concentrations around inclusions in high strength steels, Materials Science
  and Engineering: A 513 (2009) 376--383.

\bibitem{proudhon2007xray}
H.~Proudhon, J.-Y. Buffi{\`e}re, S.~Fouvry, Three-dimensional study of a
  fretting crack using synchrotron {X-ray} micro-tomography, Engineering
  Fracture Mechanics 74~(5) (2007) 782--793.

\bibitem{Alp2020}
A.~KarakoÃ§, J.~Paltakari, E.~Taciroglu, Data-driven computational
  homogenization method based on euclidean bipartite matching, Journal of
  Engineering Mechanics 146~(2) (2020) 04019132.

\bibitem{kelly2007atom}
T.~F. Kelly, M.~K. Miller, Atom probe tomography, Review of Scientific
  Instruments 78~(3) (2007) 031101.

\bibitem{spowart2006automated}
J.~E. Spowart, Automated serial sectioning for {3-D} analysis of
  microstructures, Scripta Materialia 55~(1) (2006) 5--10.

\bibitem{betz2007imaging}
O.~Betz, U.~Wegst, D.~Weide, M.~Heethoff, L.~Helfen, W.-K. Lee, P.~Cloetens,
  Imaging applications of synchrotron {X-ray} phase-contrast microtomography in
  biological morphology and biomaterials science. {I}. general aspects of the
  technique and its advantages in the analysis of millimetre-sized arthropod
  structure, Journal of Microscopy 227~(1) (2007) 51--71.

\bibitem{lecun1989}
Y.~LeCun, B.~Boser, J.~S. Denker, D.~Henderson, R.~E. Howard, W.~Hubbard, L.~D.
  Jackel, Backpropagation applied to handwritten zip code recognition, Neural
  Computation 1~(4) (1989) 541--551.

\bibitem{bishop1995neural}
C.~M. Bishop, et~al., Neural networks for pattern recognition, Oxford
  university press, 1995.

\bibitem{girosi1995regularization}
F.~Girosi, M.~Jones, T.~Poggio, Regularization theory and neural networks
  architectures, Neural Computation 7~(2) (1995) 219--269.

\bibitem{kingma2014adam}
D.~P. Kingma, J.~Ba, Adam: A method for stochastic optimization, arXiv preprint
  arXiv:1412.6980.

\bibitem{2015keras}
F.~Chollet, et~al., Keras, \url{https://github.com/fchollet/keras} (2015).

\bibitem{krizhevsky2012imagenet}
A.~Krizhevsky, I.~Sutskever, G.~E. Hinton, Imagenet classification with deep
  convolutional neural networks, in: Advances in Neural Information Processing
  Systems, 2012, pp. 1097--1105.

\bibitem{cybenko1989approximation}
G.~Cybenko, Approximation by superpositions of a sigmoidal function,
  Mathematics of Control, Signals and Systems 2~(4) (1989) 303--314.

\bibitem{du2002efficient}
X.~Du, W.~Chen, Efficient uncertainty analysis methods for multidisciplinary
  robust design, AIAA journal 40~(3) (2002) 545--552.

\bibitem{chen2006probabilistic}
C.~Chen, D.~Duhamel, C.~Soize, Probabilistic approach for model and data
  uncertainties and its experimental identification in structural dynamics:
  Case of composite sandwich panels, Journal of Sound and Vibration 294~(1-2)
  (2006) 64--81.

\bibitem{pan2009survey}
S.~J. Pan, Q.~Yang, A survey on transfer learning, IEEE Transactions on
  Knowledge and Data Engineering 22~(10) (2009) 1345--1359.

\end{thebibliography}

\end{document}